\documentclass[amsmath,superscriptaddress,longbibliography,prl,twocolumn] {revtex4-2}

\usepackage{dsfont}
\usepackage{amsmath}
\usepackage{epsfig}
\usepackage{graphicx}
\usepackage{bm}
\usepackage{amssymb}
\usepackage{slashed}
\usepackage{hyperref}
\makeatletter
\newcommand{\Rmnum}[1]{\expandafter\@slowromancap\romannumeral #1@}
\makeatletter

\begin{document}
\title{Linear magnetoconductivity in magnetic metals}
\author{Vladimir A. Zyuzin}
\affiliation{Nordita, KTH Royal Institute of Technology and Stockholm University, Roslagstullsbacken 23, SE-106 91 Stockholm, Sweden}
\affiliation{L.D. Landau Institute for Theoretical Physics, Chernogolovka, Russia}
\begin{abstract}
We theoretically describe a mechanism of low-field linear magnetoconductivity in helical magnetic metals.
Two ingredients for the mechanism in three-dimensional metals are identified to be the spin-orbit coupling and momentum-dependent ferromagnetic exchange interaction. 
We propose and study a number of minimal theoretical models which have linear magnetoconductivity, and discuss their implications for recent experiments.
\end{abstract}
\maketitle

Onsager's relations \cite{Onsager1931, MazurDeGroot1953} dictate that the low-field electric conductivity of the system in the applied magnetic field must be even under the reversal of the magnetic field when the time-reversal symmetry is not violated in the system. 
However, when the time-reversal symmetry is broken in the system by, for example, spontaneous ferromagnetic order, the Onsager relations allow for the low-field linear magnetoconductivity in the system.

\begin{figure}[t] 
\centerline{
\includegraphics[width=0.8\columnwidth,height=0.13\textheight]{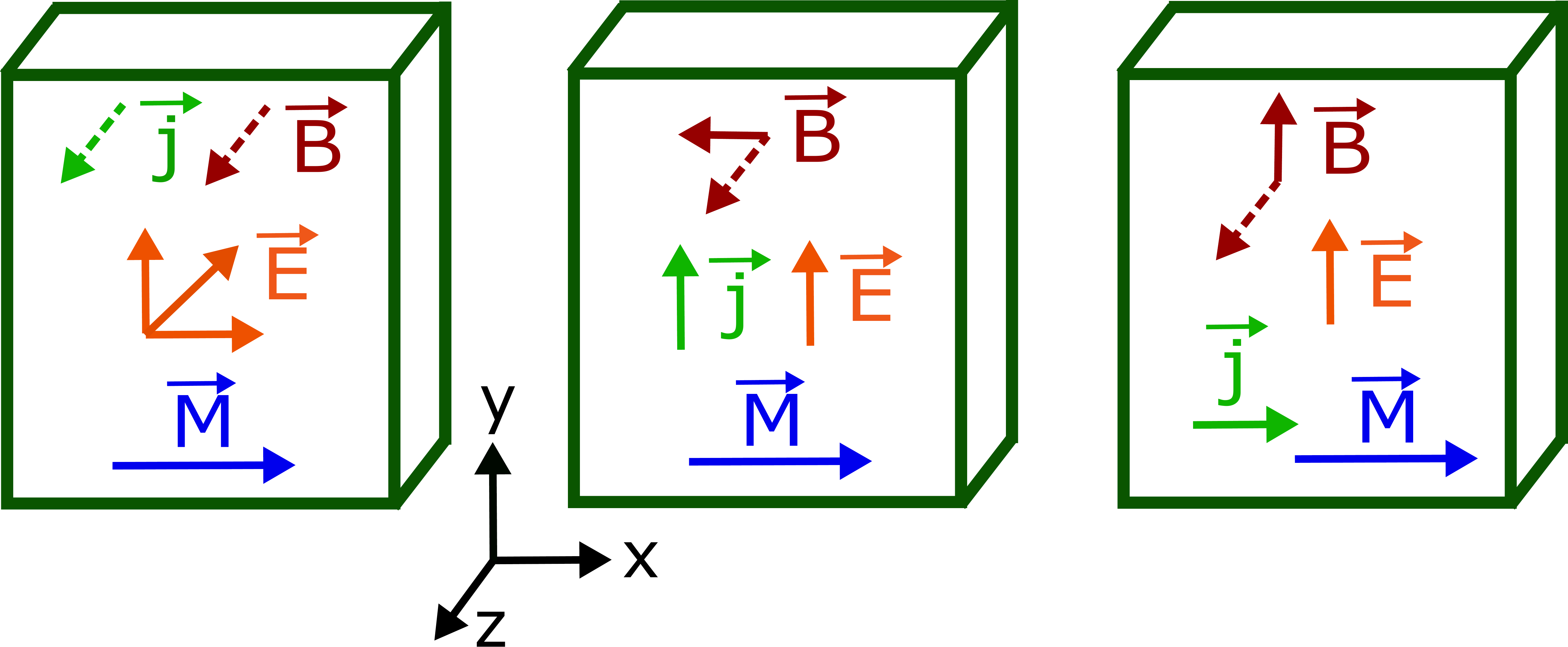}
}

\protect\caption{Schematics of the proposed experimental setup to measure three unique angular dependeces of the linear magnetoconductivity. Dashed lines are pointing in $z-$ direction. Left: the current is passed in $x-$ direction, the magnetic field is varied in $z-y$ plane, magnetization is in $x-$ direction, while the electric field is measured in the $y$ direction. Hence, only the $({\bf E}\cdot{\bf B}){\bf M}$ component of the current is active.  Center: the current is passed in $z-$ direction, magnetic field is in $z-$ direction, magnetization is in $x-$ direction, while the electric field is measured in the $x-y$ plane. In this case only the $({\bf E}\cdot{\bf M}){\bf B}$ component of the current is active. Right: the current is passed in $y-$ direction, magnetic field is varied in $z-x$ plane, magnetization is in $x-$ direction, while the electric field is measured in the $y$ direction. Thus, only the $({\bf M}\cdot{\bf B}){\bf E}$ component of the current is active.}

\label{fig1}  

\end{figure}

There is a number of recent experiments \cite{WSMCorrelated, LeeRosenbaum2020,ExpPRL2021} which observe linear magnetoconductivity in ferromagnetic metals. Indeed, based on the Onsager's relation argument, one would expect that when spontaneous magnetization ${\bf M}$ is present in the system, there might be terms in the electric current which will depend on the magnetization and result in linear magnetoconductivity. Three such possible terms with a pronounced angle dependence between the electric ${\bf E}$, magnetic ${\bf B}$ fields and magnetization are proprtional to $({\bf E}\cdot{\bf B}){\bf M}$, $({\bf E}\cdot{\bf M}){\bf B}$, and $({\bf M}\cdot{\bf B}){\bf E}$ combinations, namely,
\begin{align}\label{linearMC}
\delta{\bf j} = \alpha_{1} ({\bf E}\cdot{\bf B}){\bf M} + \alpha_{2} ({\bf E}\cdot{\bf M}){\bf B}  + \alpha_{3} ({\bf M}\cdot{\bf B}){\bf E} ,
\end{align}
where $\alpha_{1,2,3}$ are material dependent coefficients. 
Thus, varying the direction of either magnetic field, magnetization, or the current, one can identify the presence of each term in the system, see Fig. \ref{fig1}. However, besides the knowledge of the Onsager relation, the microscopic mechanism behind these three terms is still not fully understood.
The aim of the present paper is to introduce a number of theoretical models which provide a possible mechanism of linear magnetoconductivity in magnetic metals.

We assume that the spontaneous magnetization in the metals is due to the localized fermions, while the conduction fermions are responsible for the transport in these metals. 
The localized fermions interact with the conducting fermions via the ferromagnetic exchange interaction, which is proportional to the magnetization. In order to couple the magnetization with the momentum of conducting fermions we propose that the metals are helical, meaning that there is a spin-orbit coupling \cite{Dresselhaus, Vas'koBychkovRashba, SovietTI, Dyakonov} which leads to the momentum-spin locking of conducting fermions. 
In case of pure three-dimensional spin-orbit coupling, the ferromagnetic exchange interaction acting on the spin of conducting fermions just like regular Zeeman magnetic field, can't affect the velocity of fermions unless there is spin-orbit coupling affecting motion of the conducting fermions. Indeed, ferromagnetic exchange interaction acting on conducting fermions can be gauged away by simply shifting the momentum of fermions. However, we show that the momentum-dependent ferromagnetic exchange interaction \cite{Ogg1966, IvchenkoKiselevPTS1992} does affect the velocity of conduction fermions, and leads to linear magnetoconductivity with all terms present in Eq. (\ref{linearMC}). The effect of momentum dependent ferromagnetic exchange on the magnetoconductivity has already been theoretically recognized in \cite{CortijoPRB2016, ZyuzinPRB2017,CommentA}.
In case of two-dimensional spin-orbit coupling, the Zeeman-like ferromagnetic exchange interaction can affect the velocity of fermions, but only when it has a component parallel to the spin-orbit coupling vector. We discuss such a scenario in our second example of the theoretical models. We show that the current Eq. (\ref{linearMC}) will depend only on one particular component of the magnetization.

The mechanism of linear magnetoconductivity proposed in this paper is due to the effects of Berry curvature and orbital magnetization \cite{Berry,BerryReview}. The Lorentz force in all of the presented cases does not result in linear magnetoconductivity.

%-----------------------------------------------------------------------------
\textit{Three-dimensional spin-orbit coupling}.
%---------------------------------------------------------------------------
As a model of a three-dimensional metal with spin-orbit coupling (helical metal) we pick the Weyl semimetal \cite{AbrikosovBeneslavskii} with two chiralities each described by a linear spectrum. 
We also include three possible momentum-dependent terms to the Hamiltonian, which might be present due to the finite magnetization ${\bf M}$ in the system.
Our model Hamiltonian for the $s=\pm$ chiralities is
\begin{align}
\hat{H}_{s}=&
sv({\bm \sigma}\cdot{\bf k})+({\bm \sigma}\cdot{\bf M})-\mu
\label{ABC}
\\
+
& sa_{\mathrm{A}}({\bf M}\cdot{\bf k}) +a_{\mathrm{B}} \sum_{n}M_{n} k_{n}^2\sigma_{n} + a_{\mathrm{C}}({\bf M}\cdot{\bf k})({\bm \sigma}\cdot{\bf k}),
\nonumber
\end{align}
where $v$ is the velocity of conducting fermions, $\mu$ is the chemical potential, ${\bm \sigma}$ are the Pauli matrices describing spin of fermions, and ${\bf k}=(k_{x},k_{y},k_{z})$ is the three-dimensional momentum.
The term with $a_{\mathrm{A}}$ is a tilt of the Dirac cones. The tilt breaks the time-reversal symmetry. 
The other two terms in the second line of Eq. (\ref{ABC}), with $a_{\mathrm{B}}$ and $a_{\mathrm{C}}$, are momentum dependent ferromagnetic exchange interaction, which break the time-reversal symmetry as well. They were considered in \cite{Ogg1966, IvchenkoKiselevPTS1992} in studies of fermion's g-factor anisotropy in quantum wells. The second term in the first line of Eq. (\ref{ABC}) is the regular ferromagnetic exchange interaction, analogous to usual Zeeman magnetic field. This term simply splits the two chiraliries $s=\pm$ in momentum and can be shifted away from the Hamiltonian of a given chirality. Since we are interested in effects linear in ${\bf M}$ the shift will not affect the terms in the second line of Eq. (\ref{ABC}). 
In all of the models symmetry between the chiralities is broken by the terms in the second line of Eq. (\ref{ABC}).

Our three models which we will be calling as A, B, and C in accord with $a_{\mathrm{A}}$, $a_{\mathrm{B}}$, and $a_{\mathrm{C}}$ terms in Eq. \ref{ABC} correspondingly, are three dimensional metals with spin-orbit coupling. 
This implies presence of the Berry curvature and orbital magnetization in the description of the fermions \cite{Berry, BerryReview}.
To study the electric current, we employ the method of kinetic equation,
\begin{align}\label{kineticequation}
\frac{\partial n^{(s)}_{{\bf k}}}{\partial t}
+ {\dot {\bf k}}^{(s)} \frac{\partial n^{(s)}_{{\bf k}}}{\partial {\bf k}}
+ {\dot {\bf r}}^{(s)} \frac{\partial n^{(s)}_{{\bf k}}}{\partial {\bf r}}
= I_{\mathrm{coll}}[ n^{(s)}_{{\bf k}} ],
\end{align}
with equations of motion updated in the presence of the Berry curvature \cite{BerryReview},
$
{\dot {\bf r}}^{(s)} = \frac{\partial \epsilon^{(s)}_{{\bf k}} }{\partial {\bf k}}
+{\dot {\bf k}}^{(s)}\times {\bf \Omega}^{(s)}_{{\bf k}\eta},
$
and 
$
 {\dot {\bf k}}^{(s)} = e{\bf E} + \frac{e}{c}{\dot {\bf r}}^{(s)}\times{\bf B},
$
where ${\bf \Omega}^{(s)}_{{\bf k}}$ is the Berry curvature. 
The current is given by
$
{\bf j } = \sum_{s=\pm}\int_{\bf k}e\dot{{\bf r}}^{(s)}_{\bf k} n^{(s)}_{\bf k}.
$
In the fermion collision integral $I_{\mathrm{coll}}[ n^{(s)}_{{\bf k}} ]$ we consider two scattering processes described by different life-times. The first one is scattering within the chiralities denoted by $\tau$, and the other - between the chiralities denoted by $\tau_{\mathrm{V}}$, namely
$
I_{\mathrm{coll}}[ n^{(\pm)}_{{\bf k}} ]
= ({\bar n}^{(\pm)} - n^{(\pm)}_{{\bf k}})\tau^{-1}
+ ({\bar n}^{(\mp)} - n^{(\pm)}_{{\bf k}})\tau_{\mathrm{V}}^{-1},
$
where ${\bar n}^{(s)} = (4\pi)^{-1}\int \sin(\theta)d\theta d\phi [ 1 + \frac{e}{c}({\bf B}\cdot {\bf \Omega}^{(s)}_{{\bf k}})] n^{(s)}_{\bf k} $ is the distribution function averaged over the angles. 
To analyze the electric current we follow approximations used in \cite{ZyuzinPRB2017}.

In \cite{ZyuzinPRB2017} the electric current for a system with $a_{\mathrm{B}}=a_{\mathrm{C}}=0$ was studied, and it was shown that there is indeed linear magnetoconductivity due to an interplay of the chiral anomaly and the tilt of the Dirac cones and. Based on the findings of \cite{ZyuzinPRB2017} (also see Supplemental Material \cite{SM}), we here distinguish three contributions to the current. The first one is due to the chiral anomaly \cite{NielsenNinomiya}. In other words, when a difference of charges at different chirilities builds up in the presence of electric and magnetic fields, namely $N_{+}-N_{-} \propto \tau_{\mathrm{V}}({\bf E}\cdot{\bf B})$. This contribution results in $\propto\tau_{\mathrm{V}}({\bf E}\cdot{\bf B}){\bf M}$ term in the current.
Second contribution to the current is similar in nature to the first one, with the only difference being that a build up of a non-zero chiral charge in the two valleys happens for a given absolute value of the momentum when only the electric field is present. Namely, $\bar{n}^{(+)}_{k} - \bar{n}^{(-)}_{k} \propto \tau_{\mathrm{V}}({\bf E}\cdot{\bf M})$ and recall, that $N_{s}=\int \frac{k^2dk}{2\pi^2}\bar{n}^{(s)}_{k}$. We show that there is no chiral anomaly due to this contribution in all the three models. This contribution results in $\propto \tau_{\mathrm{V}}({\bf E}\cdot{\bf M}){\bf B}$ to the current.
Note that the first two contributions are defined by inter-chirality relaxation processes, and are proportional to $\tau_{\mathrm{V}}$. 
Third contribution to the current is due to the Berry curvature and orbital magnetization corrections to the fermion velocity. 
This contribution is primarily defined by relaxation processes within the chirality and, thus, defined by time $\tau$. 
All three terms present in Eq. (\ref{linearMC}) can be derived for the electric current from the third contribution. 
However, we are assuming that $\tau_{\mathrm{V}}\gg\tau$, which allows us to select from them only the unique term of the $\propto  \tau({\bf M}\cdot{\bf B}){\bf E}$ type. This assumption is legitimate given that the splitting between the chiralities defined by ${\bf M}$ is large. Details of the derivations are given in \cite{SM}.
Here we list calculated expressions for the linear magnetoransport for the three models,
\begin{widetext}
\begin{align}\label{resultA}
\delta{\bf j}_{\mathrm{A}}
\approx 
\frac{e^3 a_{\mathrm{A}} }{\pi^2 c} 
\left\{
- \tau_{\mathrm{V}}
\left[ \frac{1}{4}({\bf E}\cdot{\bf B}){\bf M} + \frac{1}{6}({\bf E}\cdot{\bf M}){\bf B} \right]
+
\frac{2\tau}{15}  ({\bf M}\cdot{\bf B}){\bf E}
\right\},
\end{align}
\begin{align}\label{resultB}
\delta{\bf j}_{\mathrm{B}}
\approx 
\frac{e^3 \mu a_{\mathrm{B}} }{ \pi^2 c v}
\left\{
- \frac{\tau_{\mathrm{V}}}{15}  
\left[ 4({\bf E}\cdot{\bf B}){\bf M} + ({\bf E}\cdot{\bf M}){\bf B} \right]
+
\frac{\tau}{7} ({\bf M}\cdot{\bf B}){\bf E} 
\right\},
\end{align}
\begin{align}\label{resultC}
\delta{\bf j}_{\mathrm{C}}
\approx 
-\frac{e^3 \mu a_{\mathrm{C}} }{ \pi^2 c v}
\left\{
\frac{\tau_{\mathrm{V}}}{9}  
\left[ ({\bf E}\cdot{\bf B}){\bf M} + ({\bf E}\cdot{\bf M}){\bf B} \right]
+ \frac{5\tau}{3}({\bf M}\cdot{\bf B}){\bf E}
\right\}. 
\end{align}
\end{widetext}
In all the three models all terms listed in Eq. (\ref{linearMC}) are present. 
The signs and numerical coefficients are model dependent. 
We also find in our calculations a $\propto ({\bf E}\cdot{\bf M})({\bf M}\cdot{\bf B}){\bf M}$ term in the current. 
However, it doesn't have a pronounced angle dependence.

%-----------------------------------------------------------------------------------------------------
\textit{Quasi two-dimensional systems.}
%----------------------------------------------------------------------------------------------------
Here we consider quasi two-dimensional system with two-dimensional Rashba spin-orbit coupling in $x-y$ plane, i.e. with Rashba spin-orbit coupling vector in $z-$ direction, and magnetization ${\bf M}$ pointing in $z-$ direction. 
This is a model of a hypotethical $\mathrm{BiTeI}$ type material with spontaneous magnetization pointing in $z-$ direction. 
The Hamiltonian of the system is
\begin{align}\label{HamD}
\hat{H}_{\mathrm{D}}  = \frac{{\bf k}^2}{2m} + \lambda(k_{x}\sigma_{y} - k_{y}\sigma_{x}) + M_{z}\sigma_{z} - \mu,
\end{align}
where ${\bf k}=(k_{x},k_{y},k_{z})$ is the three-dimensional momentum.
The spectrum consists of two branches $\epsilon_{{\bf k};\pm} = \frac{k^2}{2m} \pm \sqrt{M_{z}^2+ (\lambda k_{\parallel})^2}$, and we assume that the chemical potential $\mu$ is such that both branches are occupied.
The Berry curvature can only point in $z-$ direction. 
Moreover, integrating kinetic equation over the angles, one can check that there is no chiral anomaly in the system, meaning that the $N_{+}-N_{-} = 0$ in applied electric and magnetic fields. We approximate $\tau =\tau_{\mathrm{V}}$ as the two chiralities are close to each other in momentum and energy space.
We calculate linear magnetoconductivity following the same steps outlined above, and we get
\begin{align}
\delta {\bf j}_{\mathrm{D}}  
&
= 
\frac{e^3\lambda^2}{ 2m^2 c}\tau I_{1}
\left[ (E_{z}M_{z}) {\bf B}
+ 
 ({\bf E}\cdot{\bf B}) M_{z}{\bf e}_{z} \right]
\label{modelD}
\\
&
+ \frac{e^3\lambda^2}{8 m^2 c}\tau \left( 6I_{2} - I_{3}\right) (M_{z}B_{z}){\bf E},
\nonumber
\end{align}
where $I_{1}$, $I_{2}$, and $I_{3}$ are defined in the SM. 
Again, all terms listed in Eq. (\ref{linearMC}) are present in Eq. (\ref{modelD}), but the current depends only on $M_{z}$. 
In addition to Eq. (\ref{modelD}) we also find $\propto M_{z}B_{z}E_{z}{\bf e}_{z}$ term in the current (see \cite{SM} for details).
We can't generalize the obtained expression Eq. (\ref{modelD}) to any direction of the magnetization, because $M_{x}$ and $M_{y}$ can be shifted away from the Hamiltonian Eq. (\ref{HamD}). 
One might wonder what will happen if non-linear time-reversal symmetry breaking due to ${\bf M}=(M_{x},M_{y},M_{z})$ corrections to the Hamiltonian, similar to those with $a_{B}$ and $a_{C}$ in Eq. (\ref{ABC}), will do to the linear magnetoconductivity. 
According to \cite{ZyuzinPRB2020}, all such possible corrections which would enter the Hamiltonian Eq. (\ref{modelD}) with $\sigma_{x}$ and $\sigma_{y}$ Pauli matrices, will not affect the Berry curvature and orbital magnetization to linear order in magnetization ${\bf M}$. The result of those entering with $\sigma_{z}$ in the case when $M_{z}=0$ and $M_{x}\neq 0$ and $M_{y}\neq 0$ can be traced from the following argument. 
According to \cite{ZyuzinPRB2020}, one can think of a system which will have a non-trivial Berry curvature and orbital magnetization when $M_{z}=0$, $M_{x}\neq 0$ and $M_{y} \neq 0$ in Eq. (\ref{HamD}). 
In such case one will then need to add spin-orbit coupling term obeying, for example, a $C_{3v}$ symmetry to the Eq. (\ref{HamD}). Such a spin-orbit coupling (which can be thought of having a vector in $y$-direction) reads as $H_{\mathrm{SOC}} \propto v_{\mathrm{D}}k_{x}\sigma_{z} + \alpha k_{x}(k_{x}^2-3k_{y}^2)\sigma_{z}$, where $v_{\mathrm{D}}$ and $\alpha$ are coefficients. 
Then, even in this case, the linear magnetoconductivity will be of the Eq. (\ref{modelD}) form with the only difference of $M_{z}$ being replaced by $M_{y}$ with an appropriate coefficient. 

Finally, note that the term with $\propto ({\bf E}\cdot{\bf B})$ in Eq. (\ref{modelD}) reminds the chiral anomaly contribution, however, it is of different origin. In other words, there is no difference in chemical potentials of the two chiralities when electric and magnetic fields are applied, and as already mentioned $N_{+}-N_{-} = 0$.

%-----------------------------------------------------------------------------------------------------
\textit{Discussion.}
%-----------------------------------------------------------------------------------------------------
Typically, low-field linear magnetoconductivity has a small magnitude, and at some point it gets overshadowed by quadratic magnetoconductivity as the magnetic field is increased. 
Despite of that linear magnetoconductivity has a rich anistropic structure, which can be tested in the experiment (see Fig. \ref{fig1}). 

We think that the low-energy description of the conduction fermions in ferromagnets, which experimentally show linear magnetoconductivity, fall in to the classes of the theoretical models presented above. 
Or, it might as well be, in to some other models with the same ingredients, namely, the spin-orbit coupling and momentum dependent ferromagnetic exchange interaction. As a result, if either of the $\alpha_{1}$, $\alpha_{2}$, or $\alpha_{3}$ components of the current Eq. (\ref{linearMC}) is observed in the experiment, the other two must also be present. Based on our findings, below we comment on the two recent experiments.

In the experiment Ref. \cite{LeeRosenbaum2020} linear magnetoconductivity was observed in ferromagnetic metal SmCo$_5$ and in ferromagnetic domains of Cd$_2$Os$_2$O$_7$ antiferromagnet. 
Terms with $({\bf E}\cdot{\bf B}){\bf M}$ and $({\bf M}\cdot{\bf B}){\bf E}$ in Eq. (\ref{linearMC}) were observed in the experiment.
Based on our findings, we think that a $({\bf E}\cdot{\bf M}){\bf B}$ component of the current was overlooked \cite{CommentB}. 
We hope further experiments will identify this missing term, thus, confirming our theoretical models and discussed above mechanism of linear magnetoconductivity. Moreover, we think that the $({\bf E}\cdot{\bf B}){\bf M}$ term in the current observed in Ref. \cite{LeeRosenbaum2020} might be due to the chiral anomaly. However, further analysis should be made to eliminate possible quasi-two dimensional properties, where as is shown for model D, Eq. (\ref{modelD}), such term is present but isn't due to the chiral anomaly. 

In another experiment Ref. \cite{ExpPRL2021} linear magnetoconductivity was observed in magnetic Weyl semimetal Co$_3$Sn$_2$S$_2$ \cite{MagneticWSM2,MagneticWSM3}. There it was claimed that the effect might be due to the tilt of the Weyl cones, namely due to $a_{\mathrm{A}}$ term in the Eq. (\ref{ABC}) - mechanism first proposed in \cite{ZyuzinPRB2017}.  
Our findings introduced above suggest that the mechanism due to the tilt might not be the only one.  Below we will make one more comment on the experiment Ref. \cite{ExpPRL2021}.

Below are four comments on the model Hamiltonians Eq. (\ref{ABC}) and (\ref{HamD}).
First, any three-dimensional linear in momentum spin-orbit coupling will be described by physics of Weyl semimetals.
Hence the choice of the model Hamiltonian Eq. (\ref{ABC}) - Weyl semimetal with two chiralities. 
However, one can reengineer the Hamiltonian, for example, by adding a regular, $\propto\frac{k^2}{2m}$ type, term.
Then, in models B and C such term will allow us to simplify the spectrum by reducing it to only one valley. 
Note that the problem of chiral anomaly would not be faced in this case, and the overall charge will be conserved. 
This is because there will still be two Fermi surfaces with opposite chiralities. Model A, on the other hand, can't be reduced to only one valley.

Second, we saw that two main ingredients for the linear magnetoconductivity are the linear in momentum spin-orbit couping and momentum-dependent ferromagnetic exchange interaction. However, this is is not a unique combination, and one can achieve the same effect of momentum-dependent exchange interaction by introducing, in addition to linear spin-orbit coupling, next in expansion, if symmetry allows, cubic in momentum term. Then, ferromagnetic exchange interaction can be kept to zeroth order in momentum (just like regular Zeeman term). The two schemes are similar to each other and should result in similar linear magnetoconductivity.

Third, in the realistic bulk systems the spin-orbit might not be pure, meaning that, some Pauli matrix in the Hamiltonian is not, or only partly, describing the spin of the electrons. 
Instead, it might be describing pseudospin or some mixture of spin and, for example, unit cell's degree of freedom. 
In this case ${\bf M}$ in the components in the current will be anisotropic, and, in the most severe example, the current might only depend only on one projection of the magnetization ${\bf M}$. 
For example, in model D, Eq. (\ref{HamD}), we saw that it is only on $M_{z}$ projection the linear magnetoconductivity depends on. 
However, all three terms in the linear magnetoconductivity are present in the model.
When model D is reduced to two dimensions, only the $M_{z}B_{z}{\bf E}$ out of the three terms will survive. 

Fourth, depending on the symmetries of the crystall structure of experimental magnetic system, the magnetization ${\bf M}$ entering the current Eq. (\ref{linearMC}) and the second line in the Hamiltonian Eq. (\ref{ABC}) migth be replaced with its rotated direction, for example ${\bf M} \rightarrow {\bf M}\times {\bf e}_{x,y,z}$ or even with other configurations (for example see Eq. (3) in Ref. \cite{ZyuzinPRB2020}). Note that we have already discussed a possibility of such a replacement after Eq. (\ref{modelD}). 
Quite likely, this situation was observed in the experiment Ref. \cite{ExpPRL2021}. 
Namely, Ref. \cite{ExpPRL2021} observed two linear magnetoconductivity terms, $j_{x} \propto E_{x}M_{z}B_{y}$ and $j_{y} \propto E_{x}M_{z}B_{x}$. 
In terms of Eq. \ref{linearMC}, the two can be understood as 
\begin{align}\label{EXP}
&
\delta j_{x} = \alpha_{3} (\left[M_{z}{\bf e}_{z}\times {\bf e}_{x}\right]\cdot{\bf B}) E_{x}  ,
\\
&
\delta j_{y} = \alpha_{1} ({\bf E}\cdot{\bf B})\left[M_{z}{\bf e}_{z}\times {\bf e}_{x}\right]_{y}.
\nonumber
\end{align}
Therefore, we predict that $\delta{\bf j} = \alpha_{2}({\bf E}\cdot\left[M_{z}{\bf e}_{z}\times {\bf e}_{x}\right]){\bf B} = \alpha_{2}(E_{y}M_{z}){\bf B}$ term should also be observed in the experiment Ref. \cite{ExpPRL2021}.

Essentially, the theory presented in this paper is based on the effect of the Berry curvature on the fermion properties.
We note that there are other known scattering processes which contribute to the anomalous velocity of fermions. 
These are the skew-scattering and side-jump processes which are known, for example, to contribute to the anomalous Hall effect \cite{Smit1955, Smit1958, Berger1970}.
As is discussed in \cite{XiaoMacDonaldNiuPRB} these processes might contribute to the linear magnetoconductivity as well. 
Whether they will result in current of the Eq. (\ref{linearMC}) type is a question for future research.

Since the theory of the linear magnetoconductivity presented in this paper stems from the Berry curvature of fermions, and so as the anomalous Hall effect does too \cite{AHE_RMP}, we think that both effects, linear magnetoconductivity and the anomalous Hall effect, should be experimentally looked for in the same material.
For example, the model A is known \cite{ZyuzinTiwari} to show anomalous Hall effect as a function of the tilt, here and in \cite{ZyuzinPRB2017} we concluded that it shows linear magnetoconductivity due to the same tilt as well. 
So as the model D \cite{AHE_RMP} has the same feature. 
It can be checked that the remaining B and C models have the same property. 

In passing, more magnetic Weyl and topological semimetals have been recently experimentally identified \cite{MagneticWSM0,MagneticWSM1,MagneticWSM4}, and based on our findings here, we anticipate that linear magnetoconductivity, just like in experiments \cite{WSMCorrelated, LeeRosenbaum2020, ExpPRL2021}, should be observed in these systems. 
Moreover, we believe that linear magnetoconductivity should be added to a plethora of effects and properties such as the Fermi arcs \cite{WanTurnerVishwanathSavrasovPRB2011}, chiral anomaly driven positive longitudinal magnetoconductivity \cite{NielsenNinomiya,SonSpivak} and symmetric in magnetic field so-called planar Hall effect (both are the components of the $\delta {\bf j}\propto({\bf E}{\bf B}){\bf B}$ current \cite{ZyuzinPRB2017}, and see comment [30] in Ref. \cite{ZyuzinPRB2020}), anomalous Hall effect (due to the chirality splitting \cite{YangLuRanPRB2011} and due to the tilt $a_{\mathrm{A}}$ alone \cite{ZyuzinTiwari}), chiral collective modes \cite{ZZ2015, GorbarMiranskyShovkovySukhachovPRL2017}, and others, which make Weyl semimetals unique physical systems \cite{GorbarMiranskyShovkovySukhachovBook}.

%-----------------------------------------------------------------------------------------------------
\textit{Conclusions.}
%-----------------------------------------------------------------------------------------------------
In this letter we theoretically discussed the mechanism of linear magnetoconductivity in magnetic metals. 
We identified two necessary ingredients for the minimal model of the Hamiltonian of conducting fermionsand - three-dimensional spin-orbit coupling and momentum dependent coupling to the magnetization. If the spin-orbit coupling is two-dimensional, the coupling to the magnetization is of regular exchange interaction.  We proposed and studied four models Eq. (\ref{ABC}) and (\ref{HamD}) of such two scenarios.
In all of the models linear magnetoconductivity contains three unique terms outlined in Eq. (\ref{linearMC}), with the model dependent coefficients, see Eqs. (\ref{resultA}), (\ref{resultB}), (\ref{resultC}), and (\ref{modelD}).

\textit{Acknowledgements.}
The author is thankful to Pirinem School of Theoretical Physics, where parts of this work were completed, for hospitality.  
This work is supported by the the Russian Foundation for Basic Research (grant No. 20-52-12013) – Deutsche Forschungsgemeinschaft (grant No. EV 30/14-1) cooperation, and by the Foun-
dation for the Advancement of Theoretical Physics and
Mathematics “BASIS”.
This work is also supported by the VILLUM FONDEN via the Centre of Excellence for Dirac Materials (Grant No. 11744), the European Research Council under the European Unions Seventh Framework Program Synergy HERO, and the Knut and Alice Wallenberg Foundation.

%------------------------------------------------------------------------------------------------------
%------------------------------------------------------------------------------------------------------
%------------------------------------------------------------------------------------------------------
%------------------------------------------------------------------------------------------------------

\clearpage
\onecolumngrid
\begin{center}
\rule{0.38\linewidth}{1pt}\\
\vspace{-0.37cm}\rule{0.49\linewidth}{1pt}
\end{center}
\setcounter{section}{0}
\setcounter{equation}{0}

\section*{Supplemental Material to \texorpdfstring{\\}{}
"Linear magnetoconductivity in magnetic metals"}

\begin{widetext}

%---------------------------------------------------------------------------------------------------------------
\section{Berry curvature and orbital magnetization}
%---------------------------------------------------------------------------------------------------------------
Here we outline derivations of Berry curvature Refs. \cite{Berry, BerryReview} in the Main Text for general two-band fermion system.
We assume the system to be three-dimensional. Thus, there must necessary be two chiralities denoted by $s=\pm$ in the system.
Spin part of the Hamiltonian for $s=\pm$ chirality is
\begin{align}
H^{(s)}_{\mathrm{spin}} 
= s{\bf g}_{\bf k}\cdot{\bm \sigma} 
= s(g_{x;{\bf k}}\sigma_{x} + g_{y;{\bf k}}\sigma_{y} + g_{z;{\bf k}}\sigma_{z}).
\end{align}
The spectrum is
\begin{align}
\epsilon^{(s)}_{\pm;{\bf k}} = \pm \vert {\bf g}^{(s)}_{\bf k}\vert,
\end{align}
For $s=\pm$ wave functions are
\begin{align}
&
\Psi^{(s=+)}_{+} = \left[ \begin{array}{c} \cos\left( \frac{\theta_{\bf k}}{2}\right)e^{-i\chi_{\bf k}} \\ \sin\left( \frac{\theta_{\bf k}}{2}\right)  \end{array}\right], 
~~~
\Psi^{(s=+)}_{-} = \left[ \begin{array}{c} -\sin\left( \frac{\theta_{\bf k}}{2}\right)e^{-i\chi_{\bf k}} \\ \cos\left( \frac{\theta_{\bf k}}{2}\right)  \end{array}\right],
\\
&
\Psi^{(s=-)}_{-} = \left[ \begin{array}{c} \cos\left( \frac{\theta_{\bf k}}{2}\right)e^{-i\chi_{\bf k}} \\ \sin\left( \frac{\theta_{\bf k}}{2}\right)  \end{array}\right], 
~~~
\Psi^{(s=-)}_{+} = \left[ \begin{array}{c} -\sin\left( \frac{\theta_{\bf k}}{2}\right)e^{-i\chi_{\bf k}} \\ \cos\left( \frac{\theta_{\bf k}}{2}\right)  \end{array}\right],
\end{align}
where $g_{x;{\bf k}} + i g_{y;{\bf k}} = g_{\parallel;{\bf k}}e^{i\chi_{\bf k}}$, where $g_{\parallel;{\bf k}} = \sqrt{g_{x;{\bf k}}^2 + g_{y;{\bf k}}^2}$ and $\sin(\theta_{\bf k}) = \frac{g_{\parallel;{\bf k}}}{\vert {\bf g}_{\bf k}\vert}$. 
We note that it only appears that the wave functions obey $\Psi^{(s=+)}_{+} = \Psi^{(s=-)}_{-}$ and $\Psi^{(s=+)}_{-} = \Psi^{(s=-)}_{+}$ and, therefore, one may conclude that they don't form an orthogonal set. However, there is another pseudo-spin type degree of freedom corresponding to a chirality $s=\pm$ which makes the wave function orthogonal to each other.
Useful identities are
\begin{align}
&
\partial_{x} \Psi^{(+)}_{+} = \frac{1}{2} \Psi^{(+)}_{-}(\partial_{x}\theta_{\bf k}) - i \left[ \begin{array}{c} \cos\left( \frac{\theta_{\bf k}}{2}\right)e^{-i\chi_{\bf k}} \\ 0  \end{array}\right] (\partial_{x}\chi_{\bf k}),
\\
&
\partial_{x} \Psi^{(+)}_{-} = -\frac{1}{2} \Psi^{(+)}_{+}(\partial_{x}\theta_{\bf k}) - i \left[ \begin{array}{c} -\sin\left( \frac{\theta_{\bf k}}{2}\right)e^{-i\chi_{\bf k}} \\ 0  \end{array}\right] (\partial_{x}\chi_{\bf k}),
\end{align}
Berry curvature is
\begin{align}
\Omega^{(s)}_{z;\pm;{\bf k}}= -i
\langle \partial_{x} \Psi^{(s)}_{\pm}  \vert \partial_{y} \Psi^{(s)}_{\pm} \rangle
- \langle \partial_{y}\Psi^{(s)}_{\pm}  \vert \partial_{x} \Psi^{(s)}_{\pm} \rangle 
= 
\mp 
\frac{s}{2}\sin(\theta_{\bf k})
\left[ (\partial_{x}\chi_{\bf k})(\partial_{y}\theta_{\bf k})
-
(\partial_{y}\chi_{\bf k})(\partial_{x}\theta_{\bf k}) 
 \right]
\end{align}
Orbital magnetization is 
\begin{align}
m^{(s)}_{z;\pm;{\bf k}} =
 \frac{i}{2}\langle \partial_{x} \Psi^{(s)}_{\pm} \vert \left[\hat{H}_{s} - \epsilon^{(s)}_{{\bf k},\pm} \right] \vert \partial_{y} \Psi^{(s)}_{\pm} \rangle
 = \frac{s}{4} \sin(\theta_{\bf k}) \left[ \epsilon^{(s)}_{{\bf k},-} - \epsilon^{(s)}_{{\bf k},+} \right]
 \left[ (\partial_{x}\chi_{\bf k})(\partial_{y}\theta_{\bf k})
-
(\partial_{y}\chi_{\bf k})(\partial_{x}\theta_{\bf k}) 
 \right].
\end{align}
We can draw a relation between the Berry curvature and orbital magnetization,
\begin{align}
{\bf m}^{(s)}_{\pm;{\bf k}} = \pm \vert {\bf g}^{(s)}_{\bf k}\vert {\bf \Omega}^{(s)}_{\pm;{\bf k}}.
\end{align}
The spectrum for the conduction (assuming chemical potential $\mu>0$) band $+$ is updated by the Berry curvature and orbital magnetization (for a review see Ref. \cite{BerryReview} in the Main Text),
\begin{align}
\varepsilon^{(s)}_{+;{\bf k}} = \epsilon^{(s)}_{+;{\bf k}} - \frac{e}{c}{\bf m}^{(s)}_{+;{\bf k}}\cdot{\bf B}.
\end{align}

In most of the cases, the Berry curvature (for example, for the conduction band) can be presented as 
\begin{align}
{\bm \Omega}^{(s)}_{{\bf k};+} = \frac{{\bf f}^{(s)}_{{\bf k}}}{\vert{\bf g}_{\bf k}^{(s)}\vert^3},
\end{align}
then the velocity for the conduction band is
\begin{align}
{\bf v}^{(s)}_{{\bf k}} = {\bf v}_{0;{\bf k}}^{(s)}\left[ 1+\frac{2e}{c}({\bm \Omega}^{(s)}_{+;{\bf k}}\cdot{\bf B})\right] - \frac{e}{c}\frac{{\bm \partial}({\bf f}^{(s)}_{{\bf k}}\cdot{\bf B})}{\vert{\bf g}_{\bf k}^{(s)}\vert^2},
\end{align}
where 
\begin{align}
{\bf v}_{0;{\bf k}}^{(s)} = \frac{\partial \epsilon^{(s)}_{+;{\bf k}}}{\partial {\bf k}}.
\end{align}
The results of this section will be used below when calculating the electric current.

%--------------------------------------------------------------------------------------------------------------
\section{Linear magnetoconductivity from kinetic equation}
%--------------------------------------------------------------------------------------------------------------

\subsection{Kinetic equation}
Here and throughout the Supplemental Material we follow approximations used in Ref. \cite{ZyuzinPRB2017} in the Main Text.
We assume that $\mu >0$, such that the conduction band is described by $\epsilon^{(s)}_{+;{\bf k}}$ spectrum.
Below we will omit the $+$ index.
Kinetic equation for the conduction band is
\begin{align}\label{kinetic_eq}
\frac{\partial n^{(s)}_{{\bf k}}}{\partial t}
+ {\dot {\bf k}}^{(s)} \cdot\frac{\partial n^{(s)}_{{\bf k}}}{\partial {\bf k}}
+ {\dot {\bf r}}^{(s)}\cdot \frac{\partial n^{(s)}_{{\bf k}}}{\partial {\bf r}}
= I_{\mathrm{coll}}\left[ n^{(s)}_{{\bf k}} \right].
\end{align}
We set $\frac{\partial n^{(s)}_{{\bf k}}}{\partial {\bf r}} = 0$ and assume a steady state $\frac{\partial n^{(s)}_{{\bf k}}}{\partial t} = 0$.

\begin{align}
\label{rdotSM}
& {\dot {\bf r}}^{(s)} = \frac{1}{\Delta^{(s)}_{{\bf k}}}
\left[ {\bf v}_{\mathrm{\bf k}}^{(s)} + e {\bf E}\times{\bf \Omega}^{(s)}_{{\bf k}} + \frac{e}{c}({\bf \Omega}^{(s)}_{{\bf k}}\cdot {\bf v}_{\mathrm{\bf k}}^{(s)} ){\bf B} \right],
\\
&
\label{kdotSM}
 {\dot {\bf k}}^{(s)} = \frac{1}{\Delta^{(s)}_{{\bf k}}}
\left[e{\bf E} + \frac{e}{c}{\bf v}_{\mathrm{\bf k}}^{(s)}\times {\bf B} + \frac{e^2}{c}\left({\bf E} \cdot{\bf B} \right){\bf \Omega}^{(s)}_{{\bf k}} \right],
\end{align}
Collision integral consists of two scattering processes,
\begin{align}
I_{\mathrm{coll}}[ n^{(\pm)}_{{\bf k}} ]
= \frac{{\bar n}^{(\pm)} - n^{(\pm)}_{{\bf k}}}{\tau}
+ \frac{{\bar n}^{(\mp)} - n^{(\pm)}_{{\bf k}}}{\tau_{\mathrm{V}}},
\end{align}
where 
\begin{align}
{\bar n}^{(\pm)}  = \langle \Delta^{(\pm)}_{{\bf k}} n_{\bf k}^{(\pm)}\rangle,
\end{align}
where
$\langle ..\rangle =\int \frac{\sin(\theta)d\theta d\phi}{4\pi}(..)$ is the integration over the angles.
The first term in the collision integral is the scattering of fermions within the $s=\pm$ chirality (valley/Weyl cone), while the second term, i.e. with $\tau_{\mathrm{V}}$, is the inter-chirality scattering of fermions. The second term is important in stabilization of the chiral chemical potential - disbalance of chemical potentials of $s=\pm$ chiralities.
The collision integral can be rewritten in a more suggestive form
\begin{align}
I_{\mathrm{coll}}[ n^{(\pm)}_{{\bf k}} ]
= \frac{{\bar n}^{(\pm)} - n^{(\pm)}_{{\bf k}}}{\tau^{*}}
+ \frac{{\bar n}^{(\mp)} - {\bar n}^{(\pm)}}{\tau_{\mathrm{V}}},
\end{align}
where 
\begin{align}
\frac{1}{\tau^{*}} = \frac{1}{\tau} + \frac{1}{\tau_{\mathrm{V}}}
\end{align}
is the total inverse fermion life-time.

Electric current is 
\begin{align}\label{currentSM}
{\bf j} = e\sum_{s=\pm}\int_{\bf k} \Delta^{(s)}_{\bf k}\dot{{\bf r}}^{(s)}n^{(s)}_{\bf k}.
\end{align}
To obtain the current we will approximate the kinetic equation and find $n^{(s)}_{\bf k}$ in the lowest order in ${\bf E}$ and ${\bf B}$.

%--------------------------------------------------------------------------------------------------------------
\subsection{Chiral anomaly contribution to the current when ${\bf B} \neq 0$ and ${\bf E} \neq 0$} 
%--------------------------------------------------------------------------------------------------------------
Chiral anomaly is the disbalance of a number of fermions of opposite chiralities. 
The chiral anomaly can be theoretically obtained from averaging the kinetic equation over the angles,
\begin{align}
\left\langle \Delta^{(s)}_{{\bf k}} {\dot {\bf k}}^{(s)}\cdot \frac{\partial n^{(s)}_{{\bf k}}}{\partial {\bf k}} \right\rangle 
= \left\langle \Delta^{(s)}_{{\bf k}} I_{\mathrm{coll}}\left[ n^{(s)}_{{\bf k}} \right] \right\rangle,
\end{align}
where $\langle ..\rangle =\int \frac{\sin(\theta)d\theta d\phi}{4\pi}(..)$ is a short notation for the integration over the angles.
To the lowest order in electric and magnetic fields, the left-hand side can be rewritten as
\begin{align}\label{Lambda_SM}
\Lambda^{(s)}\equiv
\left\langle e({\bf v}^{(s)}_{{\bf k}}\cdot{\bf E}) \frac{\partial n^{(s)}_{{\bf k}}}{\partial \epsilon_{\bf k}^{s}} \right\rangle 
+
\left\langle  \frac{e^2}{c}\left({\bf E}\cdot {\bf B} \right)({\bf \Omega}^{(s)}_{{\bf k}} \cdot{\bf v}^{(s)}_{{\bf k}})\frac{\partial n^{(s)}_{{\bf k}}}{\partial \epsilon_{\bf k}^{(s)}} \right\rangle .
\end{align}
The right-hand side is
\begin{align}
\left\langle \Delta^{(\pm)}_{{\bf k}} I_{\mathrm{coll}}[ n^{(\pm)}_{{\bf k}} ]\right\rangle
=  \frac{{\bar n}^{(\mp)} - \bar{n}^{(\pm)}}{\tau_{\mathrm{V}}},
\end{align}
note that 
\begin{align}
\left\langle \frac{{\bar n}^{(\pm)} - n^{(\pm)}_{{\bf k}}}{\tau^{*}} 
\right\rangle = 0.
\end{align}
Therefore, in steady state
\begin{align}
\Lambda^{(+)} =  \frac{{\bar n}^{(-)} - \bar{n}^{(+)}}{\tau_{\mathrm{V}}}
=
\left\langle e({\bf v}^{(+)}_{{\bf k}}\cdot{\bf E}) 
\frac{\partial n^{(+)}_{{\bf k}}}{\partial \epsilon_{\bf k}^{+}} \right\rangle 
+
\left\langle  
\frac{e^2}{c}
\left({\bf E}\cdot {\bf B} \right)
({\bf \Omega}^{(+)}_{{\bf k}}\cdot {\bf v}^{(+)}_{{\bf k}})
\frac{\partial n^{(+)}_{{\bf k}}}{\partial \epsilon_{\bf k}^{(+)}} \right\rangle,
\end{align}
with the $\Lambda^{(+)} = - \Lambda^{(-)}$ property. 
If $\Lambda^{(+)} =\frac{{\bar n}^{(-)} - \bar{n}^{(+)}}{\tau_{\mathrm{V}}} \neq 0$ then there is a chiral chemical potential in the system - disbalance of chemical potentials of opposite chiralities. It is non-zero if the right-hand side is non-zero, i.e.
$\left\langle e({\bf v}^{(+)}_{{\bf k}}\cdot{\bf E}) \frac{\partial n^{(+)}_{{\bf k}}}{\partial \epsilon_{\bf k}^{+}} \right\rangle 
+
\left\langle  \frac{e^2}{c}\left({\bf E}\cdot {\bf B} \right)({\bf \Omega}^{(+)}_{{\bf k}}\cdot {\bf v}^{(+)}_{{\bf k}})\frac{\partial n^{(+)}_{{\bf k}}}{\partial \epsilon_{\bf k}^{(+)}} \right\rangle \neq 0$. 
We distinguish two different mechanism which possibly result in non-zero right-hand side. 
The first one is when $\langle  \frac{e^2}{c}\left({\bf E}\cdot {\bf B} \right)({\bf \Omega}^{(s)}_{{\bf k}}\cdot{\bf v}^{(s)}_{{\bf k}})\frac{\partial n^{(s)}_{{\bf k}}}{\partial \epsilon_{\bf k}^{(s)}} \rangle  \neq 0$ and, as a consequence, there is a disbalance of the number of particles for $s=\pm$, namely $N_{+}-N_{-} \propto ({\bf E}\cdot{\bf B}) $, where $N_{\pm} = \int_{\bf k}n_{{\bf k},\pm}$. This is called the chiral anomaly, and it contributes to the electric current unique combinations $\propto ({\bf E}\cdot{\bf B}){\bf B}$ and $\propto ({\bf E}\cdot{\bf B}){\bf M}$.
Let us demonstrate the chiral anomaly, i.e. the disbalance of the fermion densities of the opposite chiralities when ${\bf M} = 0$, ${\bf E}\neq 0$ and ${\bf B}\neq 0$,
\begin{align}
N^{(-)}-N^{(+)}
&
=
\int \frac{k^2dk}{2\pi^2} \left[{\bar n}^{(-)} - \bar{n}^{(+)} \right] 
= 
\tau_{\mathrm{V}}\int \frac{k^2dk}{2\pi^2} \left\langle  \frac{e^2}{c}\left({\bf E} \cdot{\bf B} \right)({\bf \Omega}^{(+)}_{{\bf k}}\cdot {\bf v}^{(+)}_{{\bf k}})\frac{\partial n^{(+)}_{{\bf k}}}{\partial \epsilon_{\bf k}^{(+)}} \right\rangle
\\
&
=
-\tau_{\mathrm{V}}
({\bf E}\cdot{\bf B})\frac{e^2 v^2}{12\pi^2 c}
\int kdk\left( \frac{2}{\epsilon}\frac{\partial n}{\partial \epsilon}  -  \frac{\partial^2 n}{\partial \epsilon^2}\right)
=
-\tau_{\mathrm{V}}
({\bf E}\cdot{\bf B})\frac{e^2 }{4\pi^2 c}
\int d\epsilon \frac{\partial n}{\partial \epsilon}  
\\
&
=
\tau_{\mathrm{V}}
\frac{e^2 }{4\pi^2 c}
({\bf E}\cdot{\bf B}),
\end{align}
where $\epsilon = vk$ notation was used.

The second mechanism is when $\langle e({\bf v}^{(s)}_{{\bf k}}\cdot{\bf E}) \rangle \neq 0 $, which is possible when there is an asymmetry in the velocities of opposite chiralities $s=\pm$, namely ${\bf v}^{(+)}_{{\bf k}} \neq {\bf v}^{(-)}_{{\bf k}}$. Because of that
there is a disbalance of densities of fermions for a given absolute value of the momentum $\vert {\bf k} \vert$, i.e. difference of $\langle \Delta^{(+)}_{{\bf k}}  n_{{\bf k},+} \rangle -  \langle \Delta^{(-)}_{{\bf k}}  n_{{\bf k},-} \rangle \neq 0$.
Despite of that term, in all the examples discussed below, this disbalance does not lead to the chiral anomaly.  
In other words, after integration over the amplitude of the momentum, the disbalance vanishes, i.e. $ \int k^2dk\langle \Delta^{(+)}_{{\bf k}}  n_{{\bf k},+} \rangle - \int k^2dk \langle \Delta^{(-)}_{{\bf k}}  n_{{\bf k},-} \rangle = 0$, and $N_{
+} = N_{-}$. 
In the next subsection we will explicitly show this for the three studied in the Main Text theoretical models.
However, although there is no chiral anomaly, this mechanism contributes to the electric current with a unique contribution of $\propto ({\bf E}\cdot{\bf M}){\bf B}$.

Let us show how the two discussed mechanisms of the chirality disbalances result in the contribution to the current Eq. (\ref{currentSM}).
To calculate the current we obtain the expression for the distribution function from the kinetic equation,
\begin{align}\label{n0}
n_{{\bf k}}^{(s)} \approx 
{\bar n}^{(s)} + \tau^{*}\Lambda^{(s)} 
+ \tau^{*} e({\bf v}_{\mathrm{\bf k}}^{(s)}\cdot{\bf E}) 
\frac{\partial n^{(s)}_{\bf k}}{\partial \epsilon^{(s)}_{{\bf k}}},
\end{align}
It is enough to take $n^{(s)}_{\bf k} =  {\bar n}^{(s)}$ (also assuming that $\tau_{\mathrm{V}}\gg \tau$) from Eq. (\ref{n0}). In other words, the intra-chirality scattering processes resulted in an averaged over the angles steady-state distribution of fermions within the chiralities. 
The corresponding contribution to the current reads (see Ref. \cite{ZyuzinPRB2017} in the Main Text for details), 
\begin{align}
\delta{\bf j}_{\mathrm{\Lambda}} 
&
= e \sum_{s=\pm} \int_{{\bf k}} \left[ {\bf v}_{{\bf k}}^{(s)} + \frac{e}{c}({\bm \Omega}_{\bf k}^{(s)}\cdot{\bf v}_{{\bf k}}^{(s)} ) {\bf B}\right]
 {\bar n}^{(s)}.
\end{align}
For example, schematically, if the velocity and Berry curvature contain symmetric and anti-symmetric in $s=\pm$ terms
 ${\bf v}_{\bf k}^{(s)} = {\bf v}_{\bf k}^{(0)} + s{\bf v}_{\bf k}^{(1)}$ and ${\bm \Omega}_{\bf k}^{(s)} = s {\bm \Omega}_{\bf k}^{(0)} + {\bm \Omega}_{\bf k}^{(1)}$ then, the current becomes
 \begin{align}
\delta{\bf j}_{\mathrm{\Lambda}} 
&
= e  \int_{{\bf k}} {\bf v}_{{\bf k}}^{(0)}\left[{\bar n}^{(+)} + {\bar n}^{(-)}\right]
+ \frac{e^2}{c}  \int_{{\bf k}}\left[ ({\bm \Omega}_{\bf k}^{(1)}\cdot{\bf v}_{{\bf k}}^{(0)} ) \right]{\bf B}
\left[{\bar n}^{(+)} + {\bar n}^{(-)}\right]
\\
&
+ e  \int_{{\bf k}} {\bf v}_{{\bf k}}^{(1)}\left[{\bar n}^{(+)} - {\bar n}^{(-)}\right]
+ \frac{e^2}{c}  \int_{{\bf k}}\left[ ({\bm \Omega}_{\bf k}^{(0)}\cdot{\bf v}_{{\bf k}}^{(0)} ) \right]{\bf B}
\left[{\bar n}^{(+)} - {\bar n}^{(-)}\right]
\\
&
\equiv
e  \int_{{\bf k}} {\bf v}_{{\bf k}}^{(0)}\left[{\bar n}^{(+)} + {\bar n}^{(-)}\right]
+ \frac{e^2}{c}  \int_{{\bf k}}\left[ ({\bm \Omega}_{\bf k}^{(1)}\cdot{\bf v}_{{\bf k}}^{(0)} ) \right]{\bf B}
\left[{\bar n}^{(+)} + {\bar n}^{(-)}\right]
\\
&
- e \tau_{\mathrm{V}} \int_{{\bf k}} {\bf v}_{{\bf k}}^{(1)}\Lambda^{(+)}_{k}
- \tau_{\mathrm{V}} \frac{e^2}{c}  \int_{{\bf k}}\left[ ({\bm \Omega}_{\bf k}^{(0)}\cdot{\bf v}_{{\bf k}}^{(0)} ) \right]{\bf B}
\Lambda^{(+)}_{k},
\end{align}
where $\Lambda^{(+)}_{k} = \frac{1}{\tau_{\mathrm{V}}}\left[{\bar n}^{(-)} - {\bar n}^{(+)}\right]$.
Since ${\bar n}^{(\pm)}$ depends only on the absolute value of the momentum, $\vert {\bf k}\vert$, integration over the angles $\int_{{\bf k}} {\bf v}_{{\bf k}}^{(0)}\left[{\bar n}^{(+)} + {\bar n}^{(-)}\right] = 0$ and $\frac{e^2}{c}  \int_{{\bf k}}\left[ ({\bm \Omega}_{\bf k}^{(1)}\cdot{\bf v}_{{\bf k}}^{(0)} ) \right]{\bf B}  
\left[{\bar n}^{(+)} + {\bar n}^{(-)}\right]  = 0$. 
This is because ${\bf v}_{-{\bf k}}^{(0)} = - {\bf v}_{{\bf k}}^{(0)}$ and in three-dimensions $ {\bm \Omega}_{-\bf k}^{(0)}= -  {\bm \Omega}_{\bf k}^{(0)}$.
However, ${\bf v}_{\bf k}^{(1)}$ and $ {\bm \Omega}_{\bf k}^{(1)}$  can be even in momentum. This results in
 \begin{align} \label{chiral_current_SM}
\delta{\bf j}_{\mathrm{\Lambda}} =
- e \tau_{\mathrm{V}} \int_{{\bf k}} {\bf v}_{{\bf k}}^{(1)}\Lambda^{(+)}_{k}
- \tau_{\mathrm{V}} \frac{e^2}{c}  \int_{{\bf k}}\left[ ({\bm \Omega}_{\bf k}^{(0)}\cdot{\bf v}_{{\bf k}}^{(0)} ) \right]{\bf B}
\Lambda^{(+)}_{k}.
\end{align}
We note that if two first terms would have taken from Eq. (\ref{n0}) for the distribution function, namely $n^{(s)}_{\bf k} =  {\bar n}^{(s)} + \tau^{*}\Lambda^{(s)}_{k}$, then in the expression for the current Eq. (\ref{chiral_current_SM}) we will have  $\tau_{\mathrm{V}}\frac{\tau_{\mathrm{V}} - \tau}{\tau_{\mathrm{V}} + \tau}$ instead of just $\tau_{\mathrm{V}}$. However, as we have already mentioned, in the limit $\tau_{\mathrm{V}} \gg \tau$ we can approximate $\tau_{\mathrm{V}}\frac{\tau_{\mathrm{V}} - \tau}{\tau_{\mathrm{V}} + \tau} \approx \tau_{\mathrm{V}}$.

Let us apply this formula to the three theoretical three-dimensional models discusses in the main text. 
We remind that the Hamiltonian of the models is 
\begin{align}
\hat{H}_{s}=
sv({\bm \sigma}\cdot{\bf k})+({\bm \sigma}\cdot{\bf M})-\mu
+
sa_{\mathrm{A}}({\bf M}\cdot{\bf k}) +a_{\mathrm{B}} \sum_{n}M_{n} k_{n}^2\sigma_{n} + a_{\mathrm{C}}({\bf M}\cdot{\bf k})({\bm \sigma}\cdot{\bf k}),
\nonumber
\end{align}
with the notations defined in the main text. We note that the linear magnetoconductivity for the model A was first studied in Ref. \cite{ZyuzinPRB2017} in the Main Text, and for the model B in Ref. \cite{CortijoPRB2016} in the Main Text. We assume that $v>a_{\mathrm{A},\mathrm{B},\mathrm{C}}$ which allows us to expand spectrum, velocity, and the Berry curvature in $a_{\mathrm{A},\mathrm{B},\mathrm{C}}$ parameter.

%---------------------------------%
\subsubsection{Model A}
%---------------------------------%
For the model A described by the $a_{\mathrm{A}}\neq 0$ and $a_{\mathrm{B}}= a_{\mathrm{C}}= 0$ we calculate 
\begin{align}
\Lambda^{(+)} 
=
-\frac{e}{2k}\frac{({\bf E}\cdot{\bf M})}{M}\left( 1- \frac{v^2}{M^2} + \frac{v\mu}{M^2 k} \right)
\Theta\left(k-\frac{\mu}{v+\vert M\vert} \right)
\Theta\left(\frac{\mu}{v-\vert M\vert} - k\right)
-
\frac{e^2 v^2}{6ck}({\bf E}\cdot{\bf B})\left( \frac{2}{\epsilon}\frac{\partial n}{\partial \epsilon}  -  \frac{\partial^2 n}{\partial \epsilon^2}\right),
\end{align}
where $\Theta(x)$ is the Heaviside function, and where we used $\epsilon = vk$ notation.
Velocity is calculated to be
\begin{align}
{\bf v}^{(s)}_{\bf k} = v\frac{{\bf k}}{k}\left[ 1+\frac{2e}{c}\left( {\bf B}\cdot{\bm \Omega}_{\bf k}^{(s)}\right)\right]
+ s \frac{ev}{2ck^2}{\bf B} + s{\bf M},
\end{align}
and the Berry curvature is ${\bm \Omega}_{\bf k}^{(s)} = -s\frac{{\bf k}}{2k^3}\equiv s{\bm \Omega}_{\bf k}$. In the notations introduced above, 
${\bf v}_{\bf k}^{(0)} = v\frac{\bf k}{k}$ and ${\bf v}_{\bf k}^{(1)} = v\frac{\bf k}{k}\frac{2e}{c}\left( {\bf B}\cdot{\bm \Omega}_{\bf k}\right)+  \frac{ev}{2ck^2}{\bf B} + {\bf M}$.
We then get for the chiral anomaly current
 \begin{align}
\delta{\bf j}_{\mathrm{\Lambda}}
& =
- e \tau_{\mathrm{V}} \int_{{\bf k}} {\bf v}_{{\bf k}}^{(1)}\Lambda^{(+)}_{k}
- \tau_{\mathrm{V}} \frac{e^2}{c}  \int_{{\bf k}}\left[ ({\bm \Omega}_{\bf k}^{(0)}\cdot{\bf v}_{{\bf k}}^{(0)} ) \right]{\bf B}
\Lambda^{(+)}_{k}
\\
&
\approx
-\frac{e^3 a_{\mathrm{A}} }{\pi^2 c} 
\tau_{\mathrm{V}}
\left[ \frac{1}{4}({\bf E}\cdot{\bf B}){\bf M} + \frac{1}{6}({\bf E}\cdot{\bf M}){\bf B} \right].
\end{align}

%---------------------------------%
\subsubsection{Model B}
%---------------------------------%
For the model B described by the $a_{\mathrm{B}}\neq 0$ and $a_{\mathrm{A}}= a_{\mathrm{C}}= 0$ we calculate 
\begin{align}
\Lambda^{(+)} 
=
\frac{ea_{\mathrm{B}}}{5\pi v}({\bf E}\cdot{\bf M})\left( \epsilon^2 \frac{\partial^2 n}{\partial \epsilon^2} + 4 \epsilon \frac{\partial n}{\partial \epsilon}\right)
-
\frac{e^2 v^2}{6ck}({\bf E}\cdot{\bf B})\left( \frac{2}{\epsilon}\frac{\partial n}{\partial \epsilon}  -  \frac{\partial^2 n}{\partial \epsilon^2}\right),
\end{align}
where again $\epsilon = vk$.
For example, for the ${\bf M} = M {\bf e}_{z}$ choice, we write for the chiral anomaly current 
\begin{align}
\delta{\bf j}_{\mathrm{\Lambda}}
& = 
- e\tau_{\mathrm{V}}a_{\mathrm{B}} M_{z}{\bf e}_{z}\int_{\bf k} \left(3 - \frac{k_{z}^2}{k^2}\right)\frac{k_{z}^2}{k}\Lambda^{(+)} 
+ 2\frac{ve^2}{c}\tau_{\mathrm{V}}\int_{\bf k} {\bf k} \frac{({\bf k}\cdot{\bf B})}{k^4}\Lambda^{(+)}
\\
& \approx
-\frac{e^3 \mu a_{\mathrm{B}} }{ \pi^2 c v}
 \frac{\tau_{\mathrm{V}}}{15}  
\left[ 4({\bf E}\cdot{\bf B}){\bf M} + ({\bf E}\cdot{\bf M}){\bf B} \right],
\end{align}
where we have generalized the result to any direction of the magnetization.

%---------------------------------%
\subsubsection{Model C}
%---------------------------------%
We have for the model C,
\begin{align}
\Lambda^{(+)} =
\frac{ea_{\mathrm{C}}}{3\pi v} ({\bf E}\cdot{\bf M})\left( \epsilon^2 \frac{\partial^2 n}{\partial \epsilon^2} + 4 \epsilon \frac{\partial n}{\partial \epsilon}\right)
-
\frac{e^2 v^2}{6ck}({\bf E}\cdot{\bf B})\left( \frac{2}{vk}\frac{\partial n}{\partial \epsilon}  -  \frac{\partial^2 n}{\partial \epsilon^2}\right).
\end{align}
For the sake of calculations we assume ${\bf M} = M_{z}{\bf e}_{z}$, get
$v_{z} = v\frac{k_{z}}{k} +sa_{\mathrm{C}}M_{z}k\left(1+\frac{k_{z}^2}{k^2}\right)$, and calculate the chiral anomaly current,
 \begin{align}
\delta{\bf j}_{\mathrm{\Lambda}}
& =
- e \tau_{\mathrm{V}} \int_{{\bf k}} {\bf v}_{{\bf k}}^{(1)}\Lambda^{(+)}_{k}
- \tau_{\mathrm{V}} \frac{e^2}{c}  \int_{{\bf k}}\left[ ({\bm \Omega}_{\bf k}^{(0)}\cdot{\bf v}_{{\bf k}}^{(0)} ) \right]{\bf B}
\Lambda^{(+)}_{k}
\\
&
=
- e a_{\mathrm{C}}{\bf e}_{z}M_{z}\tau_{\mathrm{V}} 
\int_{{\bf k}} k\left(1+\frac{k_{z}^2}{k^2}\right)\Lambda^{(+)}_{k}
- \tau_{\mathrm{V}} \frac{e^2}{c}  
\int_{{\bf k}}\left[ ({\bm \Omega}_{\bf k}^{(0)}\cdot{\bf v}_{{\bf k}}^{(0)} ) \right]{\bf B}
\Lambda^{(+)}_{k}
\\
&
\approx
-\frac{e^3 \mu a_{\mathrm{C}} }{ \pi^2 c v}
\frac{\tau_{\mathrm{V}}}{9}  
\left[ ({\bf E}\cdot{\bf B}){\bf M} + ({\bf E}\cdot{\bf M}){\bf B} \right],
\end{align}
where we generalized the result to any direction of the magnetization ${\bf M}$.

%---------------------------------------------------------------------------------------------
\subsection{Absence of chiral anomaly when ${\bf B} = 0$, ${\bf E} \neq 0$, and ${\bf M}\neq 0$}
%---------------------------------------------------------------------------------------------
One may wonder if there is a chiral anomaly, i.e. fermion charge disbalance in the two chiralities, when ${\bf B} = 0$, ${\bf E} \neq 0$, and ${\bf M}\neq 0$, namely, $N_{+} - N_{-} \propto \tau_{\mathrm{V}} ({\bf E}\cdot{\bf M})$.
Here we show that
although there might
be non-zero fermion distribution disbalance in the two chiralities $s=\pm$ when just an electric field is applied, namely
$\langle \Delta^{(+)}_{{\bf k}}  n_{{\bf k},+} \rangle -  \langle \Delta^{(-)}_{{\bf k}}   n_{{\bf k},-} \rangle \propto ({\bf E}\cdot{\bf M})$,
 the chiral anomaly completely vanishes for each theoretical model we are considering.
To calculate the chiral anomaly we need to integrate $\Lambda^{+}_{k}$ over the momentum.

\subsubsection{Model A}
The integral is
\begin{align}
\int \frac{k^2 dk}{2\pi^2}\Lambda^{+}_{k}
&
=
- \frac{e}{4\pi^2}\frac{({\bf E}\cdot{\bf M})}{M}\int k dk\left( 1- \frac{v^2}{M^2} + \frac{v\mu}{M^2 k} \right)
\Theta\left(k-\frac{\mu}{v+\vert M\vert} \right)
\Theta\left(\frac{\mu}{v-\vert M\vert} - k\right)
\\
&
=
- \frac{e}{4\pi^2}\frac{({\bf E}\cdot{\bf M})}{M}
\int_{\frac{\mu}{v+\vert M \vert}}^{\frac{\mu}{v-\vert M \vert}} k dk\left( 1- \frac{v^2}{M^2} + \frac{v\mu}{M^2 k} \right)
\\
&
=
- \frac{e}{4\pi^2}\frac{({\bf E}\cdot{\bf M})}{M}
\left[
 \frac{\mu^2}{2}\left( 1-\frac{v^2}{M^2} \right)\left(\frac{1}{(v-\vert M \vert)^2} - \frac{1}{(v+\vert M \vert)^2} \right)
+
\frac{v\mu^2}{M^2}\left( \frac{1}{(v-\vert M \vert)} - \frac{1}{(v+\vert M \vert)} \right)
\right] 
\\
&
= 0,
\end{align}
calculated for $v>M$. Therefore, $N_{-} - N_{+} = 0$ as claimed.

\subsubsection{Model B and C}
The integral for the model B is
\begin{align}
\int \frac{k^2 dk}{2\pi^2}\Lambda^{+}_{k}
=
\frac{ea_{\mathrm{B}}}{10\pi^3 v}({\bf E}\cdot{\bf M})
\int k^2 dk
\left( \epsilon^2 \frac{\partial^2 n}{\partial \epsilon^2} + 4 \epsilon \frac{\partial n}{\partial \epsilon}\right),
\end{align}
where $\epsilon = vk$. In order to calculate the integral, one needs to carefully integrate second derivative by parts
\begin{align}
\int \epsilon^4  \frac{\partial^2 n}{\partial \epsilon^2}  d\epsilon =  - 4 \int \epsilon^3  \frac{\partial n}{\partial \epsilon}  d\epsilon .
\end{align}
Therefore, indeed as claimed,
\begin{align}
N_{-} - N_{+} = \tau_{\mathrm{V}}\int \frac{k^2 dk}{2\pi^2}\Lambda^{+}_{k} = 0.
\end{align}
By examining the expression for $\Lambda^{+}_{k} $ for the model C, we conclude that $\int \frac{k^2 dk}{2\pi^2}\Lambda^{+}_{k} = 0$ is true also for the model C.

%---------------------------------------------------------------------------------------------
\subsection{Berry curvature and orbital magnetization contribution to the velocity}
%---------------------------------------------------------------------------------------------

Here we will use the notation for the fermion velocity introduced in the ``Berry curvature and orbital magnetization'' section, namely
\begin{align}
{\bf v}_{{\bf k}}^{(s)} = {\bf v}_{0;{\bf k}}^{(s)}\left[ 1+\frac{2e}{c}({\bm \Omega}^{(s)}_{{\bf k}}\cdot{\bf B})\right] - \frac{e}{c}\frac{{\bm \partial}({\bf f}_{\bf k}^{(s)}\cdot{\bf B})}{\vert{\bf g}_{\bf k}^{(s)}\vert^2},
\end{align}
where ${\bf v}_{0;{\bf k}}^{(s)}= \frac{\partial \epsilon_{\bf k}^{(s)}}{\partial {\bf k}}$ is the bare fermion velocity.
See Ref. \cite{BerryReviewSM} for a review of the Berry curvature effects on fermion velocity.
Recall, that we obtain the distribution function from the kinetic equation,
\begin{align}\label{n1}
n_{{\bf k}}^{(s)} \approx 
{\bar n}^{(s)} + \tau^{*}\Lambda^{(s)} 
+ \tau^{*} e({\bf v}_{\mathrm{\bf k}}^{(s)}\cdot{\bf E}) 
\frac{\partial n^{(s)}_{\bf k}}{\partial \epsilon^{(s)}_{{\bf k}}}.
\end{align}
We have already considered the effect on the electric current of the first two terms in Eq. (\ref{n1}). They have resulted in the chiral anomaly and disbalance of fermion densities of opposite chiralities.
Now, let us understand what the third terms does.
From the third term, besides regular Drude conductivity, we obtain contributions to the current which are defined by the Berry curvature and orbital magnetization,
\begin{align}\label{BO_SM}
\delta {\bf j}_{\mathrm{BO}}
&
=
\sum_{s=\pm}\frac{e^3}{c}\tau^{*} 
\int_{\bf k} {\bf v}_{0;{\bf k}}^{(s)}
 ({\bf E}\cdot{\bf v}_{0;{\bf k}}^{(s)}) 
 ({\bf m}_{\bf k}^{(s)}\cdot {\bf B})
 \frac{\partial^2 n^{(s)}_{\bf k}}{\partial (\epsilon_{\bf k}^{(s)})^2}
 -3\sum_{s=\pm}\frac{e^3}{c}\tau^{*} 
 \int_{\bf k} {\bf v}_{0;{\bf k}}^{(s)}
  ({\bf E}\cdot{\bf v}_{0;{\bf k}}^{(s)}) 
  ({\bm \Omega}_{\bf k}^{(s)}\cdot{\bf B})
  \frac{\partial n^{(s)}_{\bf k}}{\partial \epsilon_{\bf k}^{(s)}}
\\
&
-\sum_{s=\pm}\frac{e^3}{c}\tau^{*}
\int_{\bf k}{\bf B} 
({\bm \Omega}_{\bf k}^{(s)}\cdot{\bf v}_{0;{\bf k}}^{(s)} ) ({\bf E}\cdot{\bf v}_{0;{\bf k}}^{(s)})
\frac{\partial n^{(s)}_{\bf k}}{\partial \epsilon_{\bf k}^{(s)}}
-\sum_{s=\pm}\frac{e^3}{c}\tau^{*}
\int_{\bf k}
{\bf v}_{0;{\bf k}}^{(s)} ({\bm \Omega}_{\bf k}^{(s)}\cdot{\bf v}_{0;{\bf k}}^{(s)} ) 
({\bf E}\cdot{\bf B})\frac{\partial n^{(s)}_{\bf k}}{\partial \epsilon_{\bf k}^{(s)}}
\\
&
+\sum_{s=\pm}\frac{e^3}{c}\tau^{*}
\int_{\bf k} {\bf v}_{0;{\bf k}}^{(s)}
\left[ {\bf E}\cdot{\bm \partial} ({\bf f}_{\bf k}^{(s)}\cdot{\bf B})\right]
\frac{1}{\vert {\bf g}_{{\bf k}}^{(s)}\vert^2}
\frac{\partial n^{(s)}_{\bf k}}{\partial \epsilon_{\bf k}^{(s)}}
-2
\sum_{s=\pm}\frac{e^3}{c}\tau^{*}
\int_{\bf k} {\bm \partial} ({\bf f}_{\bf k}^{(s)}\cdot{\bf B})
 ({\bf E}\cdot{\bf v}_{0;{\bf k}}^{(s)}) 
\frac{1}{\vert {\bf g}_{{\bf k}}^{(s)}\vert^2}
\frac{\partial n^{(s)}_{\bf k}}{\partial \epsilon_{\bf k}^{(s)}}.
\end{align}
For models A, B, and C we extract from the expression above only the terms which result in $({\bf M}\cdot{\bf B}){\bf E}$ contribution to the current. 
Other terms result in $({\bf E}\cdot{\bf M}){\bf B}$ and $({\bf E}\cdot{\bf B}){\bf M}$ contributions which add up to the ones obtained from the chiral anomaly contribution. We ignore these terms assuming that $\tau_{\mathrm{V}} \gg \tau$.
\begin{align}
\delta {\bf j}_{\mathrm{BO};1}
&
=
\sum_{s=\pm}\frac{e^3}{c}\tau^{*} 
\int_{\bf k} {\bf v}_{0;{\bf k}}^{(s)}
 ({\bf E}\cdot{\bf v}_{0;{\bf k}}^{(s)}) 
 ({\bf m}_{\bf k}^{(s)}\cdot {\bf B})
 \frac{\partial^2 n^{(s)}_{\bf k}}{\partial (\epsilon_{\bf k}^{(s)})^2}
 -3\sum_{s=\pm}\frac{e^3}{c}\tau^{*} 
 \int_{\bf k} {\bf v}_{0;{\bf k}}^{(s)}
  ({\bf E}\cdot{\bf v}_{0;{\bf k}}^{(s)}) 
  ({\bm \Omega}_{\bf k}^{(s)}\cdot{\bf B})
  \frac{\partial n^{(s)}_{\bf k}}{\partial \epsilon_{\bf k}^{(s)}}
 \\
 &
 =
 \sum_{s=\pm}\frac{e^3}{c}\tau^{*} 
  \int_{\bf k} {\bf v}_{0;{\bf k}}^{(s)}
  ({\bf E}\cdot{\bf v}_{0;{\bf k}}^{(s)}) 
  ({\bm \Omega}_{0; \bf k}^{(s)}\cdot{\bf B})
\left[\epsilon_{\bf k}^{(s)}  \frac{\partial^2 n^{(s)}_{\bf k}}{\partial (\epsilon_{\bf k}^{(s)})^2} - 3\frac{\partial n^{(s)}_{\bf k}}{\partial \epsilon_{\bf k}^{(s)}} \right]
\\
&
+
 \sum_{s=\pm}\frac{e^3}{c}\tau^{*} 
  \int_{\bf k} {\bf v}_{0;{\bf k}}^{(s)}
  ({\bf E}\cdot{\bf v}_{0;{\bf k}}^{(s)}) 
  ({\bm \Omega}_{1;\bf k}^{(s)}\cdot{\bf B})
\left[\epsilon  \frac{\partial^2 n}{\partial \epsilon^2} - 3\frac{\partial n}{\partial \epsilon} \right],
\end{align}
where Berry curvature was presented as ${\bm \Omega}_{\bf k}^{(s)} = {\bm \Omega}_{0; \bf k}^{(s)} + {\bm \Omega}_{1;\bf k}^{(s)}$, where ${\bm \Omega}_{0; \bf k}^{(s)} = -\frac{s{\bf k}}{2k^3}$, and ${\bm \Omega}_{1;\bf k}^{(s)}$ is proportional to the first power of the momentum-dependent ferromagnetic exchange interaction. Also we have used a $\epsilon = vk$ notation, and $n(\epsilon) = (e^{\frac{\epsilon - \mu}{T}} + 1)^{-1}$.

With the details of the models outlined in the ``Details of the theoretical models'' section (below), we present the results for the contribution to the electric current, which are linear in magnetic field and magnetization.
For the model A,
\begin{align}
\delta {\bf j}_{\mathrm{BO};1}^{(\mathrm{A})} 
= \frac{2a_{\mathrm{A}}e^3\tau^{*}}{15\pi^2 c}\left[  ({\bf E}\cdot{\bf M}){\bf B}  +  ({\bf M}\cdot{\bf B}){\bf E} +  ({\bf E}\cdot{\bf B}){\bf M} \right].
\end{align}
For the model B,
\begin{align}
\delta {\bf j}_{\mathrm{BO};1}^{(\mathrm{B})} = 
\frac{a_{\mathrm{B}}e^3\tau^{*}\mu}{7\pi^2 c v} ({\bf M}\cdot{\bf B}){\bf E} 
- \frac{a_{\mathrm{B}}e^3\tau^{*}\mu}{42\pi^2 c v} \left[  ({\bf E}\cdot{\bf M}){\bf B}   +  ({\bf E}\cdot{\bf B}){\bf M} \right]
+\frac{11a_{\mathrm{B}}e^3\tau^{*}\mu}{42\pi^2 c v}  \frac{1}{\vert {\bf M}\vert^2}({\bf E}\cdot{\bf M})({\bf B}\cdot{\bf M}){\bf M}.
\end{align}
For the model C
\begin{align}
\delta {\bf j}_{\mathrm{BO};1}^{(\mathrm{C})} = - \frac{5a_{\mathrm{C}}e^3\tau^{*}\mu}{3\pi^2 c}\left[  ({\bf E}\cdot{\bf M}){\bf B}  +  ({\bf M}\cdot{\bf B}){\bf E} +  ({\bf E}\cdot{\bf B}){\bf M} \right].
\end{align}
Note the existance of a unique $\propto \frac{1}{\vert {\bf M}\vert^2}({\bf E}\cdot{\bf M})({\bf B}\cdot{\bf M}){\bf M}$ term in the current for the model B. This term also has angle dependence.

All terms from Eq. (\ref{BO_SM}) contribute to the model D. We derive them in the ``'Details of the theoretical models' section below.

%-------------------------------------------------------------------------------------------------
\subsection{Lorentz force contribution}
%-------------------------------------------------------------------------------------------------
Let us derive contribution from the Lorentz force. Within kinetic equation it can be obtained using the Zener-Jones method, namely
\begin{align}\label{n2}
n_{{\bf k}}^{(s)} \approx 
{\bar n}^{(s)} + \tau^{*}\Lambda^{(s)} 
+ \tau e({\bf v}_{\mathrm{\bf k}}^{(s)}\cdot{\bf E}) \frac{\partial n^{(s)}_{\bf k}}{\partial \epsilon^{(s)}_{{\bf k}}}
 -\frac{\tau e}{c} [{\bf v}_{\mathrm{\bf k}}^{(s)}\times{\bf B} ]\cdot \frac{\partial}{\partial{\bf k}}
\left\{{\bar n}^{(s)} + \tau\Lambda^{(s)} +\tau e ({\bf v}_{\mathrm{k}}^{(s)}\cdot{\bf E} ) \frac{\partial n^{(s)}_{\bf k}}{\partial \epsilon^{(s)}_{{\bf k}}} \right\},
\nonumber
\end{align}
The last term is the Lorentz force contribution to the distribution function. 
Since we are interested in linear magnetoconductivity, the expression in the figured brackets must be taken to zeroth order in magnetic field.
Then, since zeroth order in magnetic field part of $\Lambda^{(s)}(\vert{\bf k}\vert)$ is due to the momentum-dependent Zeeman term, to treat it, it is enough to pick from ${\bf v}_{\mathrm{\bf k}}^{(s)}$ its linear in momentum term, i.e. ${\bf v}_{\mathrm{\bf k}}^{(s)}\propto {\bf k}$. 
Then $[{\bf v}_{\mathrm{\bf k}}^{(s)}\times{\bf B} ] \cdot\frac{\partial}{\partial{\bf k}}\Lambda^{(s)}(\vert{\bf k}\vert) = 0$.
Furthermore, if we want to compose a ${\bar n}^{(+)} - {\bar n}^{(-)}$ combination in the expression for the current, velocity in the $[{\bf v}_{\mathrm{\bf k}}^{(s)}\times{\bf B} ]$ must be taken to be proportional to the momentum-dependent ferromagnetic exchange interaction. This contribution to the velocity is proportional to $s=\pm$.
Then, we will have a ${\bar n}^{(+)} - {\bar n}^{(-)} = \tau_{\mathrm{V}}\Lambda^{(+)}\propto ({\bf E}\cdot{\bf B})$ combination, as a result. Therefore, one will get second power of the external magnetic field. Therefore, the first two terms in the figure brackets can be dopped if we are interested in the linear magnetoconductivity.
We then have for the contribution to the current,
\begin{align}
\delta {\bf j}_{\mathrm{L}} = -\frac{\tau^2 e^3}{c}\sum_{s=\pm}\int_{{\bf k}}  {\bf v}_{\mathrm{\bf k}}^{(s)} \left\{ [{\bf v}_{\mathrm{\bf k}}^{(s)}\times{\bf B} ]\cdot \frac{\partial}{\partial{\bf k}} ({\bf v}_{\mathrm{\bf k}}^{(s)}\cdot{\bf E} )\right\} \frac{\partial n^{(s)}_{\bf k}}{\partial \epsilon^{(s)}_{{\bf k}}}.
\end{align}
Apart from the regular Hall conductivity which is derived from this expression, all other linear magnetoconductivity vanish due to angle integration. We, thus, have eliminated the Lorentz contribution to the linear magnetoconductivity.

%----------------------------------------------------------------------------
\section{Details of the theoretical models}
Here we outline some details of the theoretical models we introduced in the Main Text. 

%---------------------------------------------
\subsection{Model A}
%---------------------------------------------
The Hamiltonian of the model A is
\begin{align}
\hat{H}_{s}=&
sv{\bm \sigma}\cdot{\bf k}+{\bm \sigma}\cdot{\bf M}
+sa_{\mathrm{A}}{\bf M}\cdot{\bf k}.
\end{align}
The spectrum of the model is 
\begin{align}
\epsilon^{(s)}_{\pm;\bf k} = sa_{\mathrm{A}}{\bf M}\cdot{\bf k} \pm vk,
\end{align}
where, as noted in the Main Text, the ${\bm \sigma}\cdot{\bf M}$ term is dropped.
Neither the Berry curvature, which is
\begin{align}
&
\Omega_{x}^{(s)}
 =
 -\frac{s}{2}\frac{v^3 k_{x}}{\vert {\bf g}^{(s)}_{\bf k}\vert^3 } ,
~~~
\Omega_{y}^{(s)}
 =
 -\frac{s}{2}\frac{v^3 k_{y}}{\vert {\bf g}^{(s)}_{\bf k}\vert^3 } ,
~~~
 \Omega_{z}^{(s)}
 =
 -\frac{s}{2}\frac{v^3 k_{z}}{\vert {\bf g}^{(s)}_{\bf k}\vert^3 } ,
\end{align}
nor the orbital magnetization,
\begin{align}
&
m_{x}^{(s)}
 =
 -\frac{s}{2}\frac{v^3 k_{x}}{\vert {\bf g}^{(s)}_{\bf k}\vert^2 } ,
~~~
m_{y}^{(s)}
 =
 -\frac{s}{2}\frac{v^3 k_{y}}{\vert {\bf g}^{(s)}_{\bf k}\vert^2 },
 ~~~
m_{z}^{(s)}
 =
 -\frac{s}{2}\frac{v^3 k_{z}}{\vert {\bf g}^{(s)}_{\bf k}\vert^2 } ,
\end{align}
get affected by the $a_{\mathrm{A}}$ term. 
Here $\vert {\bf g}^{(s)}_{\bf k}\vert = vk$.

%---------------------------------------------
\subsection{Model B}
%---------------------------------------------
The Hamiltonian of the model B is
\begin{align}
\hat{H}_{s}=&
sv{\bm \sigma}\cdot{\bf k}+{\bm \sigma}\cdot{\bf M}
+a_{\mathrm{B}} \sum_{n}M_{n} k_{n}^2\sigma_{n}.
\end{align}
For simplicity choose $a_{\mathrm{B}} \sum_{n}M_{n} k_{n}^2\sigma_{n} = a_{\mathrm{B}} M_{z} k_{z}^2\sigma_{z}$. 
It will be straightforward to generalize final results to any direction.
We assume that $v>a_{\mathrm{B}}$, which allows us to expand fermion spectrum, velocity, and the Berry curvature in $a_{\mathrm{B}}$.
The Berry curvature is
\begin{align}
&
\Omega_{x}^{(s)}
 =
 -\frac{s}{2}\frac{v^3 k_{x}}{\vert {\bf g}^{(s)}_{\bf k}\vert^3 } - \frac{v^2 M_{z}a_{\mathrm{B}}k_{x}k_{z}}{\vert {\bf g}^{(s)}_{\bf k}\vert^3},
~~~
\Omega_{y}^{(s)}
 =
 -\frac{s}{2}\frac{v^3 k_{y}}{\vert {\bf g}^{(s)}_{\bf k}\vert^3 } - \frac{v^2 M_{z}a_{\mathrm{B}}k_{y}k_{z}}{\vert {\bf g}^{(s)}_{\bf k}\vert^3},
~~~
 \Omega_{z}^{(s)}
 =
 -\frac{s}{2}\frac{v^3 k_{z}}{\vert {\bf g}^{(s)}_{\bf k}\vert^3 } - \frac{1}{2} \frac{v^2 M_{z}a_{\mathrm{B}}k_{z}^2}{\vert {\bf g}^{(s)}_{\bf k}\vert^3},
\end{align}
where $\vert {\bf g}^{(s)}_{\bf k}\vert = \sqrt{(vk_{x} )^2 + (vk_{y} )^2
+ (vk_{z}+sa_{\mathrm{B}}M_{z}k_{z}^2 )^2}$.
The orbital magnetization is
\begin{align}
&
m_{x}^{(s)}
 =
 -\frac{s}{2}\frac{v^3 k_{x}}{\vert {\bf g}^{(s)}_{\bf k}\vert^2 } - \frac{v^2 M_{z}a_{\mathrm{B}}k_{x}k_{z}}{\vert {\bf g}^{(s)}_{\bf k}\vert^2},
~~~
m_{y}^{(s)}
 =
 -\frac{s}{2}\frac{v^3 k_{y}}{\vert {\bf g}^{(s)}_{\bf k}\vert^2 } -  \frac{v^2 M_{z}a_{\mathrm{B}}k_{y}k_{z}}{\vert {\bf g}^{(s)}_{\bf k}\vert^2},
 ~~~
m_{z}^{(s)}
 =
 -\frac{s}{2}\frac{v^3 k_{z}}{\vert {\bf g}_{\bf k}\vert^2 } - \frac{1}{2} \frac{v^2 M_{z}a_{\mathrm{B}}k_{z}^2}{\vert {\bf g}^{(s)}_{\bf k}\vert^2}.
\end{align}
Bare fermion velocity is
\begin{align}
&
v_{x} \approx v\frac{k_{x}}{k} - s a_{\mathrm{B}} M_{z}\frac{k_{x}k_{z}^3}{k^3},
~~~
v_{y} \approx v\frac{k_{y}}{k} - s a_{\mathrm{B}} M_{z}\frac{k_{y}k_{z}^3}{k^3},
~~~
v_{z} \approx v\frac{k_{z}}{k} + s a_{\mathrm{B}} M_{z} \left(3 - \frac{k_{z}^2}{k^2}\right)\frac{k_{z}^2}{k}.
\end{align}
Quantity defining the disbalance of density of fermions of opposite chiralities is  
\begin{align}
\frac{e^2}{c}
\left\langle ({\bf E}\cdot{\bf B})({\bm \Omega}_{\bf k}^{(s)}\cdot{\bf v}_{\bf k}^{(s)})\frac{\partial n^{(s)}}{\partial \epsilon_{\bf k}^{(s)}}\right\rangle
+
e
\left\langle ({\bf E}\cdot{\bf v}_{\bf k}^{(s)}) \frac{\partial n^{(s)}}{\partial \epsilon_{\bf k}^{(s)} }\right\rangle
=
s \frac{e a_{\mathrm{B}}}{5\pi v} M_{z}E_{z}\left( \epsilon^2 \frac{\partial^2 n}{\partial \epsilon^2} + 4 \epsilon \frac{\partial n}{\partial \epsilon}\right) 
+
s\frac{e^2 v}{6ck^2} ({\bf E}\cdot{\bf B}) \left[ \epsilon  \frac{\partial^2 n}{\partial \epsilon^2}  - 2 \frac{\partial n}{\partial \epsilon} \right],
\end{align}
where $\epsilon = vk$. 
 It is important to note that 
\begin{align}
\int_{0}^{\infty}  d\epsilon ~ \epsilon^2
\left( \epsilon^2 \frac{\partial^2 n}{\partial \epsilon^2} + 4 \epsilon \frac{\partial n}{\partial \epsilon}\right) 
 =0,
\end{align}
after integration by parts. Therefore, just like in the model A, there is no chiral anomaly due to the momentum dependent ferromagnetic exchange interaction in the applied electrc field.               
We then have 
\begin{align}
\Lambda^{(s)}=
s \frac{ea_{\mathrm{B}}}{5\pi v} M_{z}E_{z}\left( \epsilon^2 \frac{\partial^2 n}{\partial \epsilon^2} 
+ 4 \epsilon \frac{\partial n}{\partial \epsilon}\right)
+
s\frac{e^2 v}{6ck^2} ({\bf E}\cdot{\bf B}) \left[ \epsilon  \frac{\partial^2 n}{\partial \epsilon^2}  - 2 \frac{\partial n}{\partial \epsilon} \right].
\end{align}
The chiral anomaly current reads
\begin{align}
\delta{\bf j}_{\mathrm{\Lambda}} = 
- e\tau_{\mathrm{V}}a_{\mathrm{B}} M_{z}{\bf e}_{z}\int_{\bf k} \left(3 - \frac{k_{z}^2}{k^2}\right)\frac{k_{z}^2}{k}\Lambda^{(+)} 
+ 2\frac{ve^2}{c}\tau_{\mathrm{V}}\int_{\bf k} {\bf k} \frac{({\bf k}\cdot{\bf B})}{k^4}\Lambda^{(+)}.
\end{align}
Integral
\begin{align}
\int_{0}^{\pi}[3-\cos^2(\theta) ]\cos^2(\theta)\sin(\theta) d\theta = \frac{8}{5}
\end{align}
appears in the course of calculation.

%---------------------------------------------
\subsection{Model C}
%---------------------------------------------
Hamiltonian of the model C is
\begin{align}
\hat{H}_{s}=&
sv{\bm \sigma}\cdot{\bf k}+{\bm \sigma}\cdot{\bf M}
+ a_{\mathrm{C}}({\bf M}\cdot{\bf k})({\bm \sigma}\cdot{\bf k}).
\end{align}
Again we will assume that $ a_{\mathrm{C}}({\bf M}\cdot{\bf k})({\bm \sigma}\cdot{\bf k}) =  a_{\mathrm{C}}M_{z}k_{z}({\bm \sigma}\cdot{\bf k})$ and generalize final results to any direction of the magnetization.
The Berry curvature and the orbital magnetization are deduced from the following calculation,
\begin{align}
\sin(\theta_{\bf k})
 \left[ (\partial_{x}\chi_{\bf k})(\partial_{y}\theta_{\bf k})
-
(\partial_{y}\chi_{\bf k})(\partial_{x}\theta_{\bf k}) 
 \right]
 =
 \frac{(v+saM_{z}k_{z})^3 k_{z}}{\vert {\bf g}^{(s)}_{\bf k}\vert^3 } 
\end{align}
\begin{align}
\sin(\theta_{\bf k})
 \left[ (\partial_{y}\chi_{\bf k})(\partial_{z}\theta_{\bf k})
-
(\partial_{z}\chi_{\bf k})(\partial_{y}\theta_{\bf k}) 
 \right]
 =
 \frac{(v+saM_{z}k_{z})^3 k_{x}}{\vert {\bf g}^{(s)}_{\bf k}\vert^3 } 
\end{align}
\begin{align}
\sin(\theta_{\bf k})
 \left[ (\partial_{z}\chi_{\bf k})(\partial_{x}\theta_{\bf k})
-
(\partial_{x}\chi_{\bf k})(\partial_{z}\theta_{\bf k}) 
 \right]
 =
 \frac{(v+saM_{z}k_{z})^3 k_{y}}{\vert {\bf g}^{(s)}_{\bf k}\vert^3 }, 
\end{align}
where $\vert {\bf g}^{(s)}_{\bf k}\vert = \sqrt{ (v+sa_{\mathrm{C}}M_{z}k_{z} )^2}k$.
Then 
\begin{align}
{\bm \Omega}^{(s)}
 =
 -\frac{s}{2}\frac{{\bf k}}{k^3 } ,
 ~~~
 {\bf m}^{(s)}
 =
 -\frac{s}{2}\frac{ (v+sa_{\mathrm{C}}M_{z}k_{z} ) k_{x}}{k^2 } .
\end{align}
Bare fermion velocity is
\begin{align}
&
v_{x} = (v+sa_{\mathrm{C}}M_{z}k_{z})\frac{k_{x}}{k},
~~~
v_{y} = (v+sa_{\mathrm{C}}M_{z}k_{z})\frac{k_{y}}{k},
~~~
v_{z} = v\frac{k_{z}}{k} +sa_{\mathrm{C}}M_{z}k\left(1+\frac{k_{z}^2}{k^2}\right).
\end{align}
The integral over the angles appearing in the course of calculation is
\begin{align}
\int_{0}^{\pi} \sin(\theta)\left[ 1+\cos^{2}(\theta) \right]d\theta = \frac{8}{3}.
\end{align}

The quantity defining the chiral anomaly is
\begin{align}
e
\left\langle ({\bf E}\cdot{\bf v}_{\bf k}^{(s)}) \frac{\partial n^{(s)}}{\partial \epsilon_{\bf k}^{(s)} }\right\rangle
+
\frac{e^2}{c}
\left\langle ({\bf E}\cdot{\bf B})({\bm \Omega}_{\bf k}^{(s)}\cdot{\bf v}_{\bf k}^{(s)})\frac{\partial n^{(s)}}{\partial \epsilon_{\bf k}^{(s)}}\right\rangle
= 
s \frac{ea_{\mathrm{C}}}{3\pi v} M_{z}E_{z}\left( \epsilon^2 \frac{\partial^2 n}{\partial \epsilon^2} + 4 \epsilon \frac{\partial n}{\partial \epsilon}\right)
+
s\frac{e^2 v}{6ck^2} ({\bf E}\cdot{\bf B}) \left[ \epsilon  \frac{\partial^2 n}{\partial \epsilon^2}  - 2 \frac{\partial n}{\partial \epsilon} \right].
\end{align}
Therefore, 
\begin{align}
\Lambda^{(s)}=
s \frac{ea_{\mathrm{C}}}{3\pi v} M_{z}E_{z}\left( \epsilon^2 \frac{\partial^2 n}{\partial \epsilon^2} + 4 \epsilon \frac{\partial n}{\partial \epsilon}\right)
+
s\frac{e^2 v}{6ck^2} ({\bf E}\cdot{\bf B}) \left[ \epsilon  \frac{\partial^2 n}{\partial \epsilon^2}  - 2 \frac{\partial n}{\partial \epsilon} \right].
\end{align}

%---------------------------------------------
\subsection{Model D}
%---------------------------------------------
The Hamiltonian of the model is
\begin{align}
\hat{H}_{\mathrm{D}}  = \frac{{\bf k}^2}{2m} + \lambda(k_{x}\sigma_{y} - k_{y}\sigma_{x}) + M_{z}\sigma_{z} - \mu.
\end{align}
The spectrum of the model is obtained,
\begin{align}
\epsilon_{\bf k}^{(s)} = \frac{k_{z}^2}{2m} + \frac{k_{\parallel}^2}{2m} + s \sqrt{ (\lambda k_{\parallel})^2 + M_{z}^2}.
\end{align}
Velocities are calculated to be
\begin{align}
v_{x}^{(s)} = k_{x}\left(\frac{1}{m} +\frac{s\lambda^2}{\sqrt{(\lambda k_{\parallel})^2 + M_{z}^2}} \right), 
~~~
v_{y}^{(s)} = k_{y}\left(\frac{1}{m} +\frac{s\lambda^2}{\sqrt{(\lambda k_{\parallel})^2 + M_{z}^2}} \right), 
~~~
v_{z}^{(s)} = \frac{k_{z}}{m}.
\end{align}
Berry curvature is only along the z-direction,
\begin{align}
\Omega^{(s)}_{z;{\bf k}} = -s \frac{\lambda^2 M_{z}}{2\left[ (\lambda k_{\parallel})^2 + M_{z}^2\right]^{\frac{3}{2}}}.
\end{align}

It can be checked that there is no chiral anomaly in the model D. For instance, in the Eq. (\ref{Lambda_SM}) which defines the chiral anomaly, $({\bf \Omega}^{(s)}_{{\bf k}} \cdot{\bf v}^{(s)}_{{\bf k}}) = \Omega^{(s)}_{z;{\bf k}}\frac{k_{z}}{m}$ and the integral over the angles is going to vanish. The same applies to $({\bf v}^{(s)}_{{\bf k}} \cdot {\bf E})$ product in Eq. (\ref{Lambda_SM}). Therefore, the chiral anomaly contribution to the current vanishes. However, one can check that contributions to the current given in Eq. (\ref{BO_SM}) can be non-zero.
We calculate them to linear order in magnetic field and magnetization
\begin{align}
\delta {\bf j}_{\mathrm{D}}  
&
= 
\frac{e^3\lambda^2}{ 2m^2 c}\tau I_{1}
\left[ (E_{z}M_{z}) {\bf B}
+ 
 ({\bf E}\cdot{\bf B}) M_{z}{\bf e}_{z} \right]
+ \frac{e^3\lambda^2}{8 m^2 c}\tau \left( 6I_{2} - I_{3}\right) (M_{z}B_{z}){\bf E}
\\
&
+
\frac{e^3\lambda^2}{8 m^2 c}\tau \left( 6I_{1}-6I_{2} + I_{3}-I_{4}\right) M_{z}B_{z}E_{z}{\bf e}_{z},
\end{align} 
where the integrals are defined as
\begin{align}
&
I_{1} =  \sum_{s=\pm}s\int_{\bf k} \frac{k_{z}^2}{\left[ (\lambda k_{\parallel})^2 + M_{z}^2\right]^{\frac{3}{2}}} 
 \frac{\partial n^{(s)}}{\partial \epsilon_{\bf k}^{(s)}}
\approx 
-\frac{m}{ (2\pi)^2}
\frac{\partial}{\partial \mu} k_{\mathrm{F}}^3\int_{0}^{1}dz \frac{\sqrt{1-z}}{M_{z}^2 + (\lambda k_{\mathrm{F}})^2 z},
\\
&
I_{2} = \sum_{s=\pm}s\int_{\bf k}\frac{k_{\parallel}^2}{\left[ (\lambda k_{\parallel})^2 + M_{z}^2\right]^{\frac{3}{2}}}
 \frac{\partial n^{(s)}}{\partial \epsilon_{\bf k}^{(s)}}
\approx
-\frac{m}{ (2\pi)^2}
\frac{\partial}{\partial \mu} k_{\mathrm{F}}^2 \int_{0}^{1}dz \frac{z}{M_{z}^2 + (\lambda k_{\mathrm{F}})^2 z}\frac{1}{\sqrt{1-z}},
\\
&
I_{3} = \sum_{s=\pm}s\int_{\bf k} \frac{k_{\parallel}^2}{ (\lambda k_{\parallel})^2 + M_{z}^2 } 
 \frac{\partial^2 n^{(s)}}{\partial (\epsilon_{\bf k}^{(s)})^2}
 \approx 
 - \frac{m}{(2\pi)^2}\frac{\partial^2}{\partial \mu^2}k_{\mathrm{F}}^3
 \int_{0}^{1}dz\frac{z}{\sqrt{M_{z}^{2} + (\lambda k_{\mathrm{F}})^2 z }}\frac{1}{\sqrt{1 - z}},
 \\
&
I_{4} = \sum_{s=\pm}s\int_{\bf k} \frac{k_{z}^2}{ (\lambda k_{\parallel})^2 + M_{z}^2 } 
 \frac{\partial^2 n^{(s)}}{\partial (\epsilon_{\bf k}^{(s)})^2}
 \approx 
  - \frac{m}{(2\pi)^2}\frac{\partial^2}{\partial \mu^2 }k_{\mathrm{F}}^3
 \int_{0}^{1}dz\frac{\sqrt{1 - z}}{\sqrt{M_{z}^{2} + (\lambda k_{\mathrm{F}})^2 z }},
\end{align}
where $\mu=\frac{k_{\mathrm{F}}^2}{2m}$, $\int_{\bf k}(...) = \int\frac{d{\bf k}}{(2\pi)^3} (...)$ is the three-dimensional integral, and we set $T=0$ in the distribution function.

\end{widetext}


\begin{references}
\bibitem{CommentB} Phenomenological theory proposed in Ref. \cite{LeeRosenbaum2020} to describe the experiment misses the $({\bf E}\cdot{\bf M}){\bf B}$ term in the current. In this theory a new force, ${\bf F} = \alpha {\bf M}\times {\bf E}$ where $\alpha$ is some constant, is added to the equations of motion of conduction fermions when there is a magnetization in the system. Then, according to \cite{LeeRosenbaum2020}, this phenomenology is limited to describing only $({\bf M}\cdot{\bf B}){\bf E}$ and $({\bf E}\cdot{\bf B}){\bf M}$ terms in the current, but can't describe the $({\bf E}\cdot{\bf M}){\bf B}$ term.  
Equation for the $\dot{{\bf k}}$ indeed can have extra terms which are due to the Berry curvature (for a review see Ref. \cite{BerryReview}), and which we have used in this paper.
\end{references}
\end{document}